\documentclass[useAMS]{mn2e}

\usepackage{graphicx}
\newcommand{\vsini}{\mbox{$v \sin I$}}

\newcommand{\mactrb}{\mbox{$v_{\rm mac}$}}

\newcommand{\etal}{et\,al.}
\newcommand{\kmps}{km\,s$^{-1}$}
\newcommand{\jup}{$_{\rm Jup}$}

\def\sqiglt{\hbox{\rlap{\lower.55ex \hbox {$\sim$}}
        \kern-.3em \raise.4ex \hbox{$<$}\,}}
\def\sqiggt{\hbox{\rlap{\lower.55ex \hbox {$\sim$}}
        \kern-.3em \raise.4ex \hbox{$>$}\,}}

\title[WASP-South hot Jupiters]{Seven transiting hot-Jupiters from WASP-South, Euler and TRAPPIST:
WASP-47b, WASP-55b, WASP-61b, WASP-62b, WASP-63b, WASP-66b \&\ WASP-67b}

\author[Hellier et al.]{Coel Hellier$^{1}$, 
D.R. Anderson$^{1}$, 
A. Collier Cameron$^{2}$, 
A. P. Doyle$^{1}$,  \newauthor  
M. Gillon$^{3}$,
E. Jehin$^{3}$, 
M. Lendl$^{4}$, 
P.F.L. Maxted$^{1}$, 
F. Pepe$^{4}$, 
D. Pollacco$^{5}$, \newauthor  
D. Queloz$^{4}$, 
D. S\'egransan$^{4}$, 
B. Smalley$^{1}$, 
A.M.S. Smith$^{1}$, 
J. Southworth$^{1}$, \newauthor 
A.H.M.J. Triaud$^{4}$, 
S. Udry$^{4}$ \&\ 
R.G. West$^{6}$\\    
$^{1}$Astrophysics Group, Keele University, Staffordshire, ST5 5BG, UK\\
$^{2}$SUPA, School of Physics and Astronomy, University of St.\ Andrews, North Haugh,  Fife, KY16 9SS, UK\\
$^{3}$Institut d'Astrophysique et de G\'eophysique, Universit\'e de
Li\`ege, All\'ee du 6 Ao\^ut, 17, Bat. B5C, Li\`ege 1, Belgium\\
$^{4}$Observatoire astronomique de l'Universit\'e de Gen\`eve
51 ch. des Maillettes, 1290 Sauverny, Switzerland\\
$^{5}$Astrophysics Research Centre, School of Mathematics \& Physics, Queen's University, University Road, Belfast, BT7 1NN, UK\\
$^{6}$Department of Physics and Astronomy, University of Leicester, Leicester, LE1 7RH, UK}

\begin{document}

\date{date}
\pagerange{range}

\maketitle

\begin{abstract}
We present seven new transiting hot Jupiters from the WASP-South
survey.  The planets are all typical hot Jupiters orbiting stars from
F4 to K0 with magnitudes of $V$ = 10.3 to 12.5. The orbital periods
are all in the range 3.9--4.6 d, the planetary masses range from
0.4--2.3 M\jup\ and the radii from 1.1--1.4 M\jup.  In line
with known hot Jupiters, the planetary densities range from
Jupiter-like to inflated ($\rho$ = 0.13--1.07 $\rho_{\rm Jup}$).
We use the
increasing numbers of known hot Jupiters to investigate the
distribution of their orbital periods and the 3--4-d ``pile-up''. 
\end{abstract}


\begin{table}
\caption{Observations\protect\rule[-1.5mm]{0mm}{2mm}}  
\begin{tabular}{lcr}
\hline 
Facility & Date &  \\ [0.5mm] \hline
\multicolumn{3}{l}{{\bf WASP-47:}}\\  
WASP-South & 2008 Jun--2010 Oct & 18\,300 points \\
Euler/CORALIE  & 2010 May--2011 Nov  &   19 radial velocities \\
EulerCAM  & 2011 Aug 02 & Gunn r filter \\ [0.5mm] 
\multicolumn{3}{l}{{\bf WASP-55:}}\\  
WASP-South & 2006 May--2010 Jul & 28\,200 points \\
Euler/CORALIE  & 2011 Jan--2011 Jul  &   20 radial velocities \\
TRAPPIST & 2011 Apr 04 & $I+z$ filter \\
EulerCAM  & 2011 Jun 03 & Gunn $r$ filter \\ 
TRAPPIST & 2012 Jan 20 & $I+z$ filter \\ [0.5mm] 
\multicolumn{3}{l}{\bf WASP-61:}\\  
WASP-South & 2006 Sep--2010 Feb & 30\,700 points \\
Euler/CORALIE  & 2011 Jan--2011 Nov  &   15 radial velocities \\
TRAPPIST & 2011 Sep 09 & Blue-block filter \\
EulerCAM  & 2011 Nov 16 & Gunn $r$ filter \\ 
TRAPPIST & 2011 Nov 16 & Blue-block filter \\ 
TRAPPIST & 2011 Dec 09 & Blue-block filter \\ 
TRAPPIST & 2011 Dec 13  & Blue-block filter \\ [0.5mm] 
\multicolumn{3}{l}{\bf WASP-62:}\\  
WASP-South & 2008 Sep--2011 Feb  & 21\,700 points \\
Euler/CORALIE  & 2011 Mar--2012 Apr  &   25 radial velocities \\
EulerCAM  & 2011 Nov 24 & Gunn $r$ filter \\ 
TRAPPIST & 2011 Dec 17 & $z$-band filter \\ [0.5mm] 
\multicolumn{3}{l}{\bf WASP-63:}\\  
WASP-South & 2006 Oct--2010 Mar & 24\,700 points \\
Euler/CORALIE  & 2011 Feb--2012 Apr  &   23 radial velocities \\
TRAPPIST & 2011 Dec 04 & $z$-band filter \\
EulerCAM  & 2011 Dec 25 & Gunn $r$ filter \\ 
EulerCAM  & 2012 Jan 29 & Gunn $r$ filter \\ 
TRAPPIST & 2012 Feb 21  & $I+z$ filter \\ [0.5mm] 
\multicolumn{3}{l}{\bf WASP-66:}\\  
WASP-South & 2006 May--2011 Jun & 19\,600 points \\
Euler/CORALIE  & 2011 Jan--2012 Mar  &   30 radial velocities \\
TRAPPIST & 2011 Apr 08 & $I+z$ filter \\
TRAPPIST & 2011 Dec 21 & $I+z$ filter \\
TRAPPIST & 2012 Mar 16 & Blue-block filter \\
EulerCAM  & 2012 Mar 16 & Gunn $r$ filter \\ [0.5mm] 
\multicolumn{3}{l}{\bf WASP-67:}\\ [0.5mm] 
WASP-South & 2006 May--2010 Sep & 12\,500 points \\
Euler/CORALIE  & 2011 Jul--2011 Oct  &   19 radial velocities \\
TRAPPIST & 2011 Sep 29 & $I+z$ filter \\
EulerCAM  & 2011 Sep 29 & Gunn $r$ filter \\ [0.5mm] \hline
\end{tabular} 
\end{table} 

\section{Introduction}
Transiting exoplanets found by the ground-based transit
searches are mostly ``hot Jupiters'', Jupiter-sized planets
in $\approx$\,1--6-d orbits, since these are the easiest planets
for such surveys to find.   However, planet candidate lists
from {\sl Kepler\/} show that hot Jupiters are much less common
than smaller planets (Batalha \etal\ 2012).   
This means that the much larger sky
coverage of the ground-based surveys (e.g.\ WASP, Pollacco \etal\ 2006,
and HATnet, Bakos \etal\ 2004) is needed to produce large samples
of hot Jupiters that will enable us to understand the
properties of this class.  In addition, hot Jupiters from
these surveys orbit stars of $V$ $\approx$ 9--13, which
are bright enough for radial-velocity measurements of the
planetary masses and for many other types of study.  

Here we present seven new transiting planets discovered by the
WASP-South survey (Hellier \etal\ 2011a) in conjunction with the
Euler/CORALIE spectrograph and the TRAPPIST robotic photometer (Jehin 
\etal\ 2011).  These are all hot Jupiters with $\sim$\,4-d orbits that
are compatible with being circular, and with masses and radii that
are typical of the class.  They all orbit relatively isolated stars of
$V$ = 10.3--12.5 which have metallicities and space velocities
compatible with local thin-disc stars. WASP-South planets that are
less typical of the class will be reported in other papers.

\section{Observations}
WASP-South uses an 8-camera array that covers 
450 square degrees of sky observing with a typical 
cadence of 8 mins.
The WASP surveys are described in Pollacco \etal\ (2006) 
while a discussion of our planet-hunting methods can 
be found in Collier-Cameron \etal\ (2007a) and Pollacco \etal\ (2007).

WASP-South planet candidates are followed up using the
TRAPPIST robotic photometer and the CORALIE spectrograph on 
the Euler 1.2-m telescope at La Silla.  About 1 in 12 candidates 
turns out to be a planet,
the remainder being blends that are unresolved in
the WASP data (which uses 14'' pixels) or astrophysical
transit mimics, usually eclipsing binary stars.   A list
of observations reported in this paper is given in Table~1
while the CORALIE radial velocities are listed in Table~A1.

\section{The host stars} 
The CORALIE spectra of the host stars were co-added to produce
spectra for analysis using the methods
described in Gillon et\,al.\ (2009).  We used the H$\alpha$ line to
determine the effective temperature ($T_{\rm eff}$), and the Na\,{\sc
i}\,D and Mg\,{\sc i}\,b lines as diagnostics of the surface gravity
($\log g_{\rm *}$). The resulting parameters are listed in Tables~2 to 8.
The
elemental abundances were determined from equivalent-width
measurements of several clean and unblended lines. A value for
microturbulence ($\xi_{\rm t}$) was determined from Fe\,{\sc i} 
lines using 
the criteria of a null-dependence of line abundances with equivalent 
width (see Magain 1984). The quoted error estimates include that given
by the uncertainties in $T_{\rm eff}$, $\log g_{\rm *}$ and $\xi_{\rm t}$, as
well as the scatter due to measurement and atomic data uncertainties.

The projected stellar rotation velocities (\vsini) were determined by
fitting the profiles of several unblended Fe~{\sc i} lines. We
used values for macroturbulence (\mactrb) from the tabulation by 
Bruntt \etal\ (2010). A CORALIE 
instrumental FWHM of 0.11 $\pm$ 0.01~{\AA} was determined from the
telluric lines around 6300\AA. 

\subsection{Rotational modulation}
We searched the WASP photometry of each star for rotational
modulations by using a sine-wave fitting algorithm as described
by Maxted \etal\ (2011). We estimated the
significance of periodicities by subtracting the fitted transit lightcurve
and then repeatedly and randomly permuting the nights of observation.  
For none of our stars was a significant periodicity obtained,
with 95\%-confidence upper limits being typically 1 mmag 
(as listed in Tables~2 to 8).

\subsection{Proper motions} 
For each of our stars we list (Tables~2 to 8) the proper
motions from the UCAC3 catalogue (Zacharias \etal\ 2010). 
Combining these with the spectroscopic distances and the 
radial velocities in the same Tables gives space velocities
in the range 19--55  \kmps.  Our stars
are all compatible with the local thin-disc population, which 
typically has $-0.6 < {\rm [Fe/H]} < 0.3$
and $\sigma_{v} \approx 50$ \kmps\ (Navarro \etal\ 2011) .

\section{System parameters}
The CORALIE radial-velocity measurements were combined with the WASP,
EulerCAM and TRAPPIST photometry in a simultaneous Markov-chain
Monte-Carlo (MCMC) analysis to find the system parameters.
For details of our methods see Collier Cameron \etal\ (2007b). 
The limb-darkening parameters are noted in each Table, and are
taken from the 4-parameter non-linear
law of Claret (2000).

For all of our planets the data are compatible with zero eccentricity 
and hence we imposed a circular orbit (see Anderson \etal\
2012 for a discussion of the rationale for this). 
The fitted parameters were thus $T_{\rm c}$,
$P$, $\Delta F$, $T_{14}$, $b$, $K_{\rm 1}$, where $T_{\rm c}$ is the
epoch of mid-transit, $P$ is the orbital period, $\Delta F$ is the
fractional flux-deficit that would be observed during transit in the
absence of limb-darkening, $T_{14}$ is the total transit duration
(from first to fourth contact), $b$ is the impact parameter of the
planet's path across the stellar disc, and $K_{\rm 1}$ is the stellar
reflex velocity semi-amplitude.

The transit lightcurves lead directly to stellar density but one additional constraint is required
to obtain stellar masses and radii, and hence full parametrisation of the
system. We adopt the approach of Enoch \etal\ (2010), based on
empirical calibrations of stellar properties from well-studied
detached eclipsing binary systems, but we use the calibration
coefficients calculated by Southworth (2011). These are
improvements on the coefficients from Enoch \etal\ as
they include far more stars (180 versus 38) and also restrict the
calibration sample to objects with masses relevant to the study of
transiting planetary systems (mass $<$\,3\,M$_{\odot}$). The rms scatter of
the calibrating sample around the best fit is 0.027\,dex for log(mass)
and 0.009\,dex for log(radius).

\newpage

\begin{figure}
\hspace*{-5mm}\includegraphics[width=10cm]{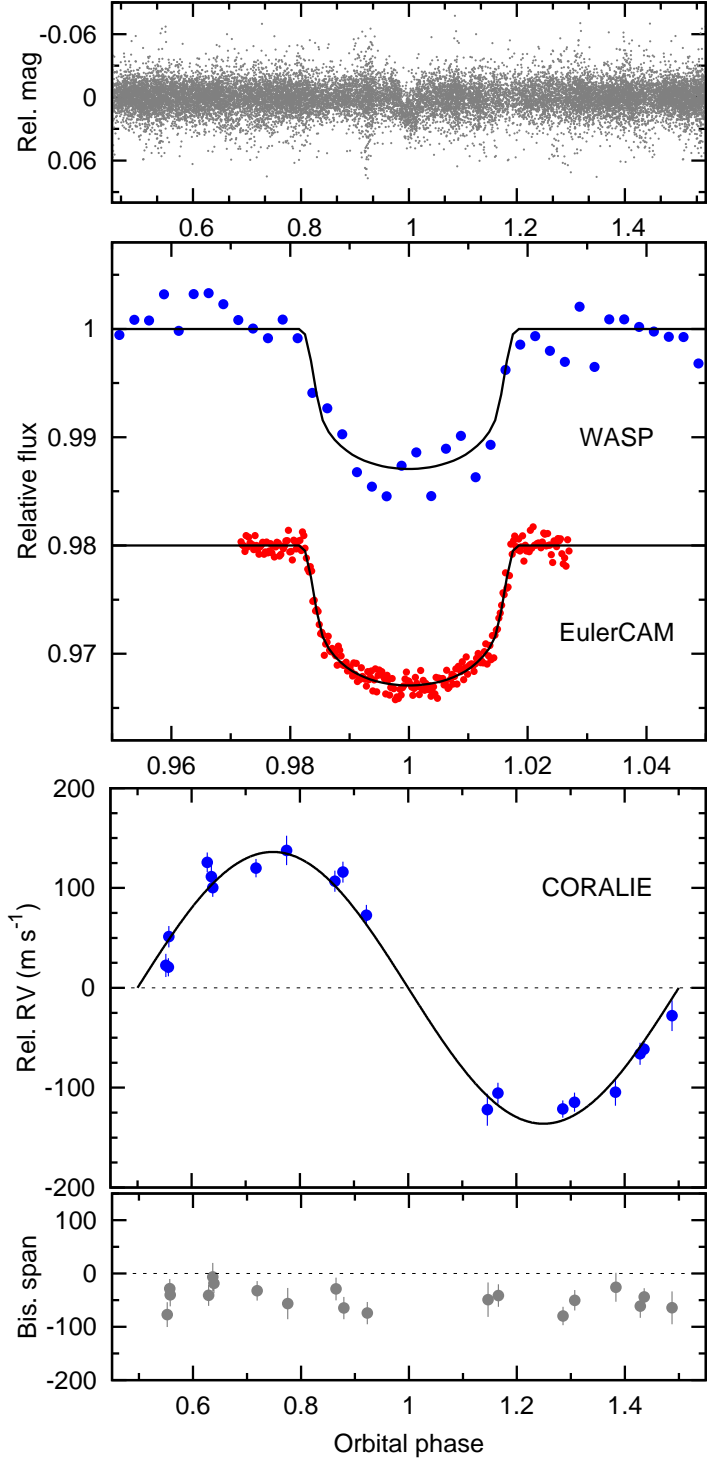}\\ [-2mm]
\caption{WASP-47b discovery data: (Top) The WASP-South lightcurve folded on the transit
period. (Second panel) The binned WASP data with (offset) the
follow-up transit lightcurves (ordered from the top as 
in Table~1) together with the 
fitted MCMC model.  (Third) The CORALIE radial
velocities with the fitted model.
(Lowest) The bisector spans; the absence of any correlation with
radial velocity is a check against transit mimics.}
\end{figure}

\begin{table}
\caption{System parameters for WASP-47.}  
\begin{tabular}{lc}
\multicolumn{2}{l}{1SWASP\,J220448.72--120107.8}\\
\multicolumn{2}{l}{2MASS 22044873--1201079}\\
\multicolumn{2}{l}{RA\,=\,22$^{\rm h}$04$^{\rm m}$48.72$^{\rm s}$, 
Dec\,=\,--12$^{\circ}$01$^{'}$07.8$^{''}$ (J2000)}\\
\multicolumn{2}{l}{$V$ mag = 11.9}  \\ 
\multicolumn{2}{l}{Rotational modulation\ \ \ $<$\,0.7 mmag (95\%)}\\
\multicolumn{2}{l}{pm (RA) 17.1\,$\pm$\,1.1 (Dec) --42.9\,$\pm$\,1.0 mas/yr}\\
\hline
\multicolumn{2}{l}{Stellar parameters from spectroscopic analysis.\rule[-1.5mm]{0mm}{2mm}} \\ \hline 
Spectral type & G9V \\
$T_{\rm eff}$ (K)      & 5400 $\pm$ 100  \\
$\log g$      & 4.55 $\pm$ 0.10 \\
$\xi_{\rm t}$ (km\,s$^{-1}$)    & 0.7 $\pm$ 0.2 \\ 
$v\,\sin I$ (km\,s$^{-1}$)     & 3.0 $\pm$ 0.6 \\
{[Fe/H]}   &   0.18 $\pm$ 0.07 \\
{[Na/H]}   &   0.42 $\pm$ 0.06 \\ 
{[Mg/H]}   &   0.21 $\pm$ 0.04 \\
{[Si/H]}   &   0.36 $\pm$ 0.07 \\
{[Ca/H]}   &   0.15 $\pm$ 0.11 \\
{[Ti/H]}   &   0.28 $\pm$ 0.06 \\
{[Cr/H]}   &   0.21 $\pm$ 0.10 \\ 
{[Ni/H]}   &   0.30 $\pm$ 0.09 \\
log A(Li)  &   $<$ 0.81 $\pm$ 0.10 \\
Distance   &   200 $\pm$ 30 pc \\ [0.5mm] \hline
\multicolumn{2}{l}{Parameters from MCMC analysis.\rule[-1.5mm]{0mm}{3mm}} \\
\hline 
$P$ (d) & 4.1591399 $\pm$ 0.0000072 \\
$T_{\rm c}$ (HJD)\,(UTC) & 2455764.34602 $\pm$ 0.00022\\ 
$T_{\rm 14}$ (d) & 0.14933 $\pm$ 0.00065\\ 
$T_{\rm 12}=T_{\rm 34}$ (d) & 0.0141$^{+ 0.0007}_{- 0.0003}$\\
$\Delta F=R_{\rm P}^{2}$/R$_{*}^{2}$ & 0.01051 $\pm$ 0.00014\\ 
$b$ & 0.14 $\pm$ 0.11\\
$i$ ($^\circ$) & 89.2 $^{+ 0.5}_{- 0.7}$\\
$K_{\rm 1}$ (km s$^{-1}$) & 0.136 $\pm$ 0.005\\ 
$\gamma$ (km s$^{-1}$) & --27.056 $\pm$ 0.004\\ 
$e$ & 0 (adopted) ($<$0.11 at 3$\sigma$) \\ 
$M_{\rm *}$ (M$_{\rm \odot}$) & 1.084 $\pm$ 0.037\\ 
$R_{\rm *}$ (R$_{\rm \odot}$) & 1.15 $^{+ 0.03}_{- 0.02}$\\
$\log g_{*}$ (cgs) & 4.348$^{+ 0.009}_{- 0.016}$\\
$\rho_{\rm *}$ ($\rho_{\rm \odot}$) & 0.71$^{+ 0.02}_{- 0.04}$\\
$T_{\rm eff}$ (K) & 5350 $\pm$ 90\\
$M_{\rm P}$ (M$_{\rm Jup}$) & 1.14 $\pm$ 0.05\\
$R_{\rm P}$ (R$_{\rm Jup}$) & 1.15 $^{+ 0.04}_{- 0.02}$\\
$\log g_{\rm P}$ (cgs) & 3.29$^{+ 0.02}_{- 0.03}$\\
$\rho_{\rm P}$ ($\rho_{\rm J}$) & 0.74$^{+ 0.05}_{- 0.06}$\\
$a$ (AU)  & 0.0520 $\pm$ 0.0006\\
$T_{\rm P, A=0}$ (K) & 1220 $\pm$ 20\\ [0.5mm] \hline 

\multicolumn{2}{l}{Errors are 1$\sigma$; Limb-darkening coefficients were:}\\
\multicolumn{2}{l}{(Euler $r$) a1 =    0.727, a2 = --0.653, a3 =  1.314, 
a4 = --0.594}\\ \hline
\end{tabular} 
\end{table}

For each system we list the resulting parameters in Tables~2 to 8,
and plot the resulting data and models in Figures~1 to 7. 
We also refer the reader to Smith \etal\ (2012) who present 
an extensive analysis of the effect of red noise in the transit
lightcurves on the resulting system parameters.  

As in past WASP papers we plot the spectroscopic $T_{\rm eff}$, and 
the stellar density from fitting the transit, 
against the evolutionary tracks
from Girardi \etal\ (2000), as shown in Fig.~8.

\newpage

\rule{0mm}{1cm}\\
\rule{0mm}{1cm}\\
\rule{0mm}{1cm}\\

\begin{figure}
\hspace*{-5mm}\includegraphics[width=10cm]{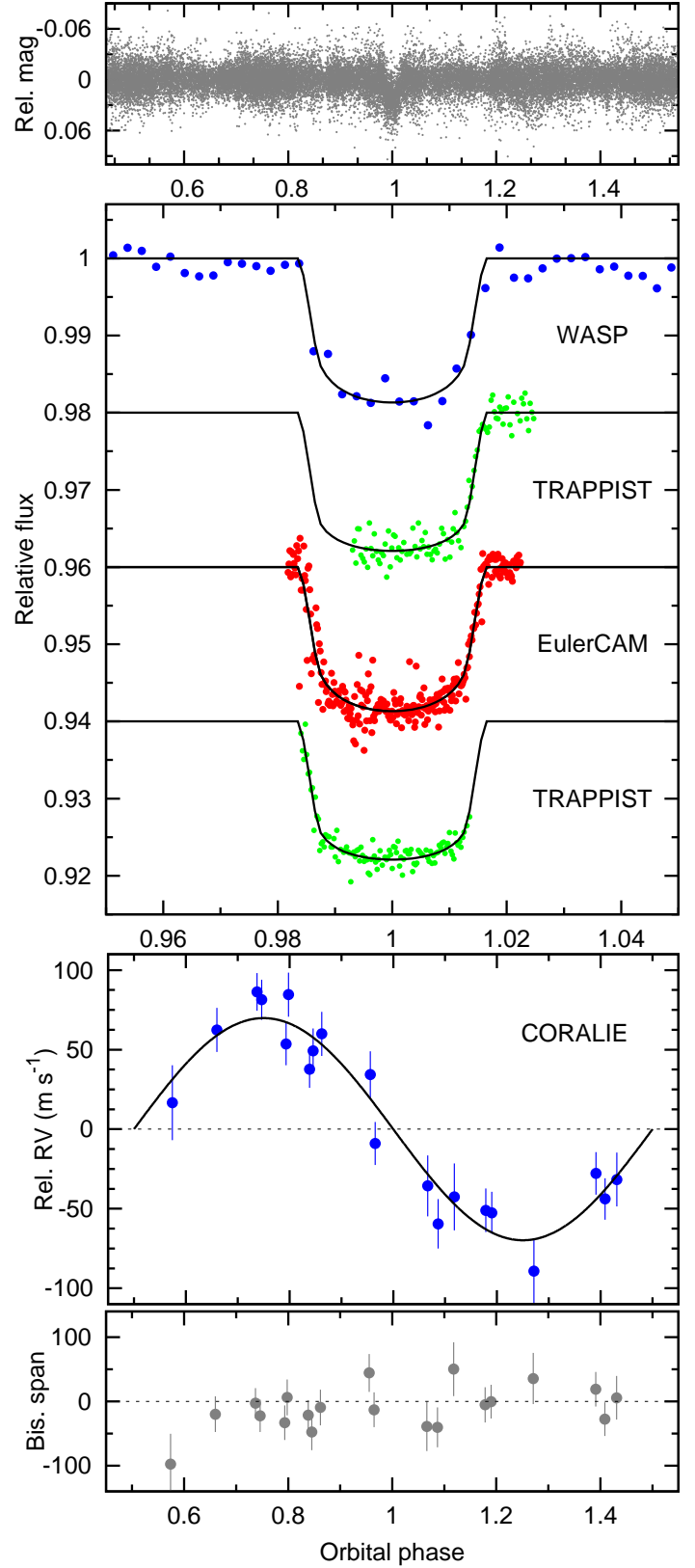}\\ [-2mm]
\caption{WASP-55b discovery data (as in Fig.~1).}
\end{figure}

\begin{table} 
\caption{System parameters for WASP-55.}  
\begin{tabular}{lc}
\multicolumn{2}{l}{1SWASP\,J133501.94--173012.7}\\
\multicolumn{2}{l}{2MASS\,13350194--1730124}\\
\multicolumn{2}{l}{TYCHO-2 6125-113-1}\\
\multicolumn{2}{l}{
RA\,=\,13$^{\rm h}$35$^{\rm m}$01.94$^{\rm s}$, 
Dec\,=\,--17$^{\circ}$30$^{'}$12.7$^{''}$ (J2000)}\\
\multicolumn{2}{l}{$V$ mag = 11.8}  \\ 
\multicolumn{2}{l}{Rotational modulation\ \ \ $<$\,1 mmag (95\%)}\\
\multicolumn{2}{l}{pm (RA) 11.2\,$\pm$\,1.0 (Dec) --8.2\,$\pm$\,1.0 mas/yr}\\
\hline
\multicolumn{2}{l}{Stellar parameters from spectroscopic analysis.\rule[-1.5mm]{0mm}{2mm}} \\ \hline 
Spectral type & G1 \\
$T_{\rm eff}$ (K)      & 5900 $\pm$ 100  \\
$\log g$      & 4.3 $\pm$ 0.1 \\
$\xi_{\rm t}$ (km\,s$^{-1}$)    & 1.1 $\pm$ 0.1 \\ 
$v\,\sin I$ (km\,s$^{-1}$)     & 3.1 $\pm$ 1.0 \\
{[Fe/H]}   &$-$0.20 $\pm$ 0.08 \\
{[Na/H]}   &$-$0.21 $\pm$ 0.05 \\ 
{[Mg/H]}   &$-$0.17 $\pm$ 0.04 \\
{[Si/H]}   &$-$0.13 $\pm$ 0.05 \\
{[Ca/H]}   &$-$0.10 $\pm$ 0.10 \\
{[Sc/H]}   &$-$0.05 $\pm$ 0.08 \\
{[Ti/H]}   &$-$0.08 $\pm$ 0.05 \\
{[Cr/H]}   &$-$0.18 $\pm$ 0.07 \\  
{[Ni/H]}   &$-$0.21 $\pm$ 0.06 \\
$\log A$(Li)  &   2.36 $\pm$ 0.09 \\
Distance   & 330 $\pm$ 50 pc  \\ [0.5mm] \hline
\multicolumn{2}{l}{Parameters from MCMC analysis.\rule[-1.5mm]{0mm}{3mm}} \\
\hline 
$P$ (d) & 4.465633 $\pm$ 0.000004 \\ 
$T_{\rm c}$ (HJD) (UTC)  & 2455737.9396 $\pm$ 0.0003 \\ 
$T_{\rm 14}$ (d) & 0.147 $\pm$ 0.001 \\ 
$T_{\rm 12}=T_{\rm 34}$ (d) & 0.0167$^{+ 0.0011}_{- 0.0004}$\\
$\Delta F=R_{\rm P}^{2}$/R$_{*}^{2}$ & 0.0158 $\pm$ 0.0003 \\
$b$ & 0.15 $\pm$ 0.12 \\ 
$i$ ($^\circ$) & 89.2 $\pm$ 0.6\\
$K_{\rm 1}$ (km s$^{-1}$) & 0.070 $\pm$ 0.004 \\
$\gamma$ (km s$^{-1}$)  & --4.3244 $\pm$ 0.0009 \\
$e$ & 0 (adopted) ($<$0.20 at 3$\sigma$) \\ 
$M_{\rm *}$ (M$_{\rm \odot}$) & 1.01 $\pm$ 0.04 \\
$R_{\rm *}$ (R$_{\rm \odot}$) & 1.06$^{+ 0.03}_{- 0.02}$\\
$\log g_{*}$ (cgs) & 4.39$^{+ 0.01}_{- 0.02}$\\
$\rho_{\rm *}$ ($\rho_{\rm \odot}$) & 0.85$^{+ 0.03}_{- 0.07}$\\
$T_{\rm eff}$ (K) & 5960 $\pm$ 100\\
$M_{\rm P}$ (M$_{\rm Jup}$) & 0.57 $\pm$ 0.04\\ 
$R_{\rm P}$ (R$_{\rm Jup}$) & 1.30$^{+ 0.05}_{- 0.03}$\\
$\log g_{\rm P}$ (cgs) & 2.89 $\pm$ 0.04\\
$\rho_{\rm P}$ ($\rho_{\rm J}$) & 0.26$^{+ 0.02}_{- 0.03}$\\
$a$ (AU)  & 0.0533 $\pm$ 0.0007 \\ 
$T_{\rm P, A=0}$ (K) & 1290 $\pm$ 25 \\ [0.5mm] \hline 
\multicolumn{2}{l}{Errors are 1$\sigma$; Limb-darkening coefficients were:}\\
\multicolumn{2}{l}{(Euler $r$) a1 =    0.496, a2 = 0.201, a3 =  0.183, 
a4 = --0.170}\\
\multicolumn{2}{l}{(Trapp $Iz$) a1 = 0.587, a2 = --0.180, a3 =  
0.441, a4 = --0.250}\\ \hline
\end{tabular} 
\end{table}

\newpage 

\rule{0mm}{1cm}\\
\rule{0mm}{1cm}\\
\rule{0mm}{1cm}\\

\begin{figure}
\hspace*{-5mm}\includegraphics[width=10cm]{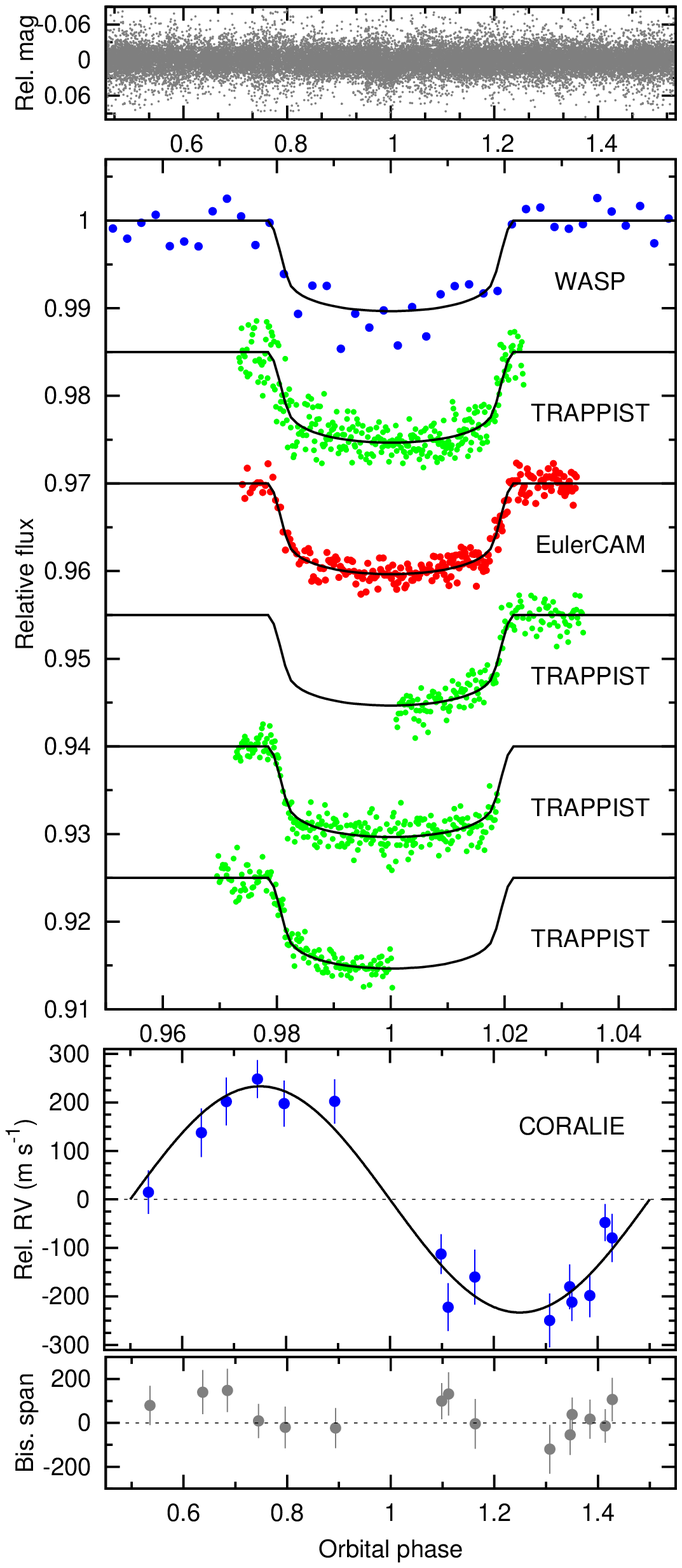}\\ [-2mm]
\caption{WASP-61b discovery data (as in Fig.~1).}
\end{figure}

\begin{table} 
\caption{System parameters for WASP-61.}
\begin{tabular}{lc}
\multicolumn{2}{l}{1SWASP\,J050111.91--260314.9}\\
\multicolumn{2}{l}{2MASS\,05011191--2603149}\\
\multicolumn{2}{l}{TYCHO-2 6469-1972-1}\\
\multicolumn{2}{l}{
RA\,=\,05$^{\rm h}$01$^{\rm m}$11.91$^{\rm s}$, 
Dec\,=\,--26$^{\circ}$03$^{'}$14.9$^{''}$ (J2000)}\\
\multicolumn{2}{l}{$V$ mag = 12.5}  \\
\multicolumn{2}{l}{Rotational modulation\ \ \ $<$\,1.5 mmag (95\%)}\\
\multicolumn{2}{l}{pm (RA) 1.0\,$\pm$\,0.9 (Dec) 0.5\,$\pm$\,1.0 mas/yr}\\
\hline
\multicolumn{2}{l}{Stellar parameters from spectroscopic analysis.\rule[-1.5mm]{0mm}{2mm}} \\ \hline 
Spectral type & F7 \\ 
$T_{\rm eff}$ (K)      & 6250 $\pm$ 150  \\
$\log g$      & 4.3 $\pm$ 0.1 \\
$\xi_{\rm t}$ (km\,s$^{-1}$)    & 1.0 $\pm$ 0.2 \\ 
$v\,\sin I$ (km\,s$^{-1}$)     & 10.3 $\pm$ 0.5 \\
{[Fe/H]}   &   --0.10 $\pm$ 0.12 \\
log A(Li)  &   $<$ 1.13 $\pm$ 0.11 \\
Distance   &   480 $\pm$ 65 pc \\ [0.5mm] \hline
\multicolumn{2}{l}{Parameters from MCMC analysis.\rule[-1.5mm]{0mm}{3mm}} \\
\hline 
$P$ (d) & 3.855900 $\pm$ 0.000003 \\
$T_{\rm c}$ (HJD) (UTC) & 2455859.52825 $\pm$ 0.00023 \\
$T_{\rm 14}$ (d) & 0.1642 $\pm$ 0.0006 \\
$T_{\rm 12}=T_{\rm 34}$ (d) & 0.0142$^{+ 0.0004}_{- 0.0002}$\\
$\Delta F=R_{\rm P}^{2}$/R$_{*}^{2}$ & 0.0088 $\pm$ 0.0001 \\
$b$ & 0.09$^{+ 0.09}_{- 0.06}$\\
$i$ ($^\circ$)  & 89.35$^{+ 0.45}_{- 0.66}$\\
$K_{\rm 1}$ (km s$^{-1}$) & 0.233 $\pm$ 0.016 \\
$\gamma$ (km s$^{-1}$) & 18.970 $\pm$ 0.002\\
$e$ & 0 (adopted) ($<$0.26 at 3$\sigma$) \\ 
$M_{\rm *}$ (M$_{\rm \odot}$) & 1.22 $\pm$ 0.07 \\
$R_{\rm *}$ (R$_{\rm \odot}$) & 1.36 $\pm$ 0.03 \\
$\log g_{*}$ (cgs) & 4.256 $\pm$ 0.011 \\
$\rho_{\rm *}$ ($\rho_{\rm \odot}$) & 0.487$^{+ 0.008}_{- 0.017}$\\
$T_{\rm eff}$ (K) & 6320 $\pm$ 140\\
$M_{\rm P}$ (M$_{\rm Jup}$) & 2.06 $\pm$ 0.17 \\
$R_{\rm P}$ (R$_{\rm Jup}$) & 1.24 $\pm$ 0.03 \\
$\log g_{\rm P}$ (cgs) & 3.48 $\pm$ 0.03 \\
$\rho_{\rm P}$ ($\rho_{\rm J}$) & 1.07 $\pm$ 0.09\\
$a$ (AU)  & 0.0514 $\pm$ 0.0009 \\
$T_{\rm P, A=0}$ (K) & 1565 $\pm$ 35 \\ [0.5mm] \hline 
\multicolumn{2}{l}{Errors are 1$\sigma$; Limb-darkening coefficients were:}\\
\multicolumn{2}{l}{(All) a1 = 0.466, a2 = 0.414, a3 =  --0.192, 
a4 = 0.002}\\ \hline
\end{tabular} 
\end{table}

\newpage 

\rule{0mm}{1cm}\\
\rule{0mm}{1cm}\\

\begin{figure}
\hspace*{-5mm}\includegraphics[width=10cm]{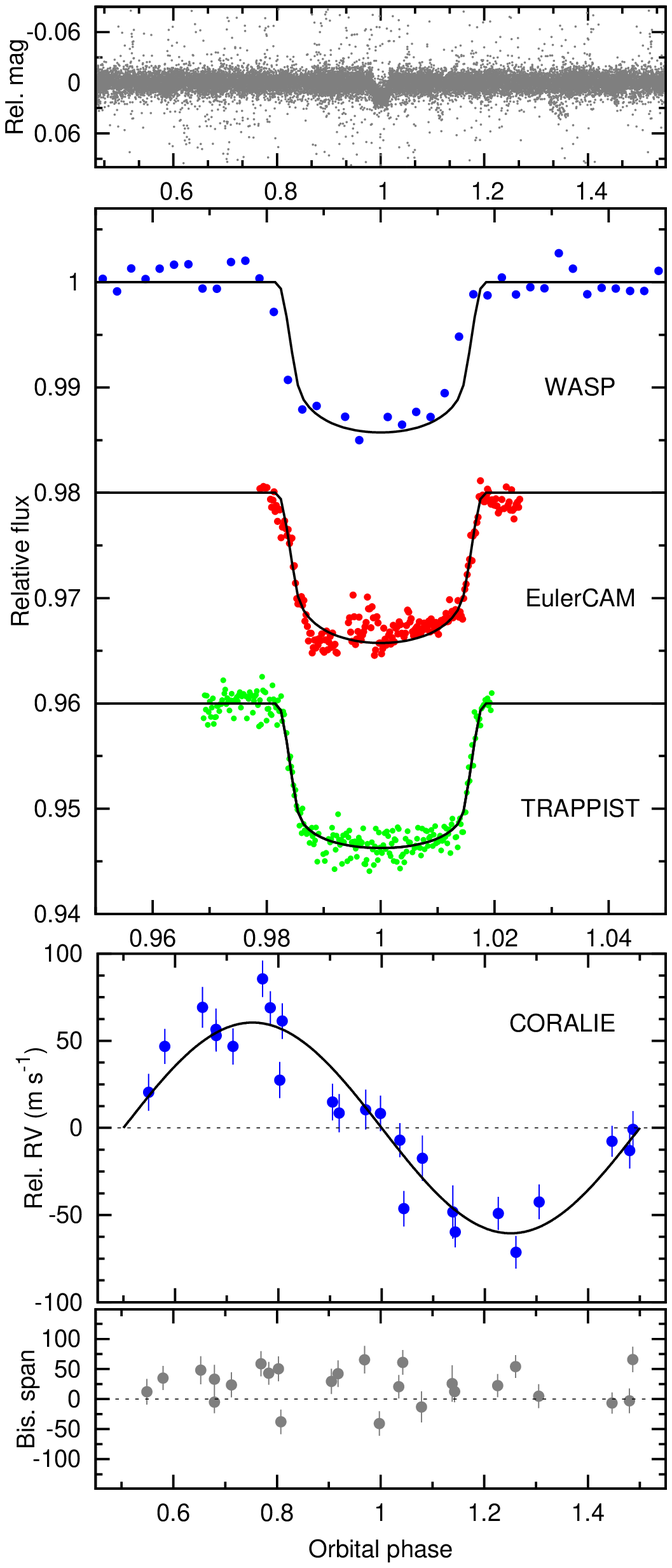}\\ [-2mm]
\caption{WASP-62b discovery data (as in Fig.~1).}
\end{figure}

\begin{table} 
\caption{System parameters for WASP-62.}
\begin{tabular}{lc}
\multicolumn{2}{l}{1SWASP\,J054833.59--635918.3}\\
\multicolumn{2}{l}{2MASS\,05483359--6359183}\\
\multicolumn{2}{l}{TYCHO-2 8900-874-1}\\
\multicolumn{2}{l}{
RA\,=\,05$^{\rm h}$48$^{\rm m}$33.59$^{\rm s}$, 
Dec\,=\,--63$^{\circ}$59$^{'}$18.3$^{''}$ (J2000)}\\
\multicolumn{2}{l}{$V$ mag = 10.3}  \\ 
\multicolumn{2}{l}{Rotational modulation\ \ \ $<$\,1 mmag (95\%)}\\
\multicolumn{2}{l}{pm (RA) --14.0\,$\pm$\,0.9 (Dec) --27.0\,$\pm$\,1.0 mas/yr}\\
\hline
\multicolumn{2}{l}{Stellar parameters from spectroscopic analysis.\rule[-1.5mm]{0mm}{2mm}} \\ \hline 
Spectral type & F7 \\
$T_{\rm eff}$ (K)      & 6230 $\pm$ 80  \\
$\log g$      & 4.45 $\pm$ 0.10 \\
$\xi_{\rm t}$ (km\,s$^{-1}$)    & 1.25 $\pm$ 0.10 \\ 
$v\,\sin I$ (km\,s$^{-1}$)     & 8.7 $\pm$ 0.4 \\
{[Fe/H]}   &   0.04 $\pm$ 0.06 \\
{[Na/H]}   &$-$0.02 $\pm$ 0.03 \\ 
{[Mg/H]}   &   0.07 $\pm$ 0.08 \\
{[Al/H]}   &   0.03 $\pm$ 0.03 \\
{[Si/H]}   &   0.11 $\pm$ 0.08 \\
{[Ca/H]}   &   0.16 $\pm$ 0.12 \\
{[Sc/H]}   &   0.10 $\pm$ 0.05 \\
{[Ti/H]}   &   0.11 $\pm$ 0.08 \\ 
{[V/H]}    &   0.01 $\pm$ 0.09 \\
{[Cr/H]}   &   0.09 $\pm$ 0.07 \\ 
{[Mn/H]}   &$-$0.08 $\pm$ 0.05 \\
{[Co/H]}   &$-$0.02 $\pm$ 0.10 \\
{[Ni/H]}   &   0.04 $\pm$ 0.08 \\
$\log A$(Li)  &   2.48 $\pm$ 0.06 \\
Distance   & 160 $\pm$ 30 pc  \\ [0.5mm] \hline
\multicolumn{2}{l}{Parameters from MCMC analysis.\rule[-1.5mm]{0mm}{3mm}} \\
\hline 
$P$ (d) & 4.411953  $\pm$ 0.000003 \\
$T_{\rm c}$ (HJD) (UTC) & 2455855.39195 $\pm$ 0.00027 \\
$T_{\rm 14}$ (d) & 0.1588 $\pm$ 0.0014 \\
$T_{\rm 12}=T_{\rm 34}$ (d) & 0.0172 $\pm$ 0.0012 \\
$\Delta F=R_{\rm P}^{2}$/R$_{*}^{2}$ & 0.0123 $\pm$ 0.0002 \\
$b$ & 0.29$^{+ 0.08}_{- 0.14}$\\
$i$ ($^\circ$)  & 88.3$^{+ 0.9}_{- 0.6}$\\
$K_{\rm 1}$ (km s$^{-1}$) & 0.060 $\pm$ 0.004 \\
$\gamma$ (km s$^{-1}$) & 14.970 $\pm$ 0.005\\
$e$ & 0 (adopted) ($<$ 0.21 at 3$\sigma$)\\ 
$M_{\rm *}$ (M$_{\rm \odot}$) & 1.25 $\pm$ 0.05 \\
$R_{\rm *}$ (R$_{\rm \odot}$) & 1.28 $\pm$ 0.05 \\
$\log g_{*}$ (cgs) & 4.316 $\pm$ 0.025 \\
$\rho_{\rm *}$ ($\rho_{\rm \odot}$) & 0.59 $\pm$0.06\\
$T_{\rm eff}$ (K) & 6280 $\pm$ 80\\
$M_{\rm P}$ (M$_{\rm Jup}$) & 0.57 $\pm$ 0.04 \\
$R_{\rm P}$ (R$_{\rm Jup}$) & 1.39 $\pm$ 0.06 \\
$\log g_{\rm P}$ (cgs) & 2.83 $\pm$ 0.04 \\
$\rho_{\rm P}$ ($\rho_{\rm J}$) & 0.21 $\pm$ 0.03\\
$a$ (AU)  & 0.0567 $\pm$ 0.0007 \\
$T_{\rm P, A=0}$ (K) & 1440 $\pm$ 30 \\ [0.5mm] \hline 
\multicolumn{2}{l}{Errors are 1$\sigma$; Limb-darkening coefficients were:}\\
\multicolumn{2}{l}{(Euler $r$) a1 =    0.508, a2 = 0.269, a3 =  0.015, 
a4 = --0.090}\\
\multicolumn{2}{l}{(Trapp $Iz$) a1 = 0.585, a2 = --0.095, a3 =  
0.276, a4 = --0.176}\\ \hline
\end{tabular}
\end{table}

\newpage

\rule{0mm}{1cm}\\
\rule{0mm}{1cm}\\

\begin{figure}
\hspace*{-5mm}\includegraphics[width=10cm]{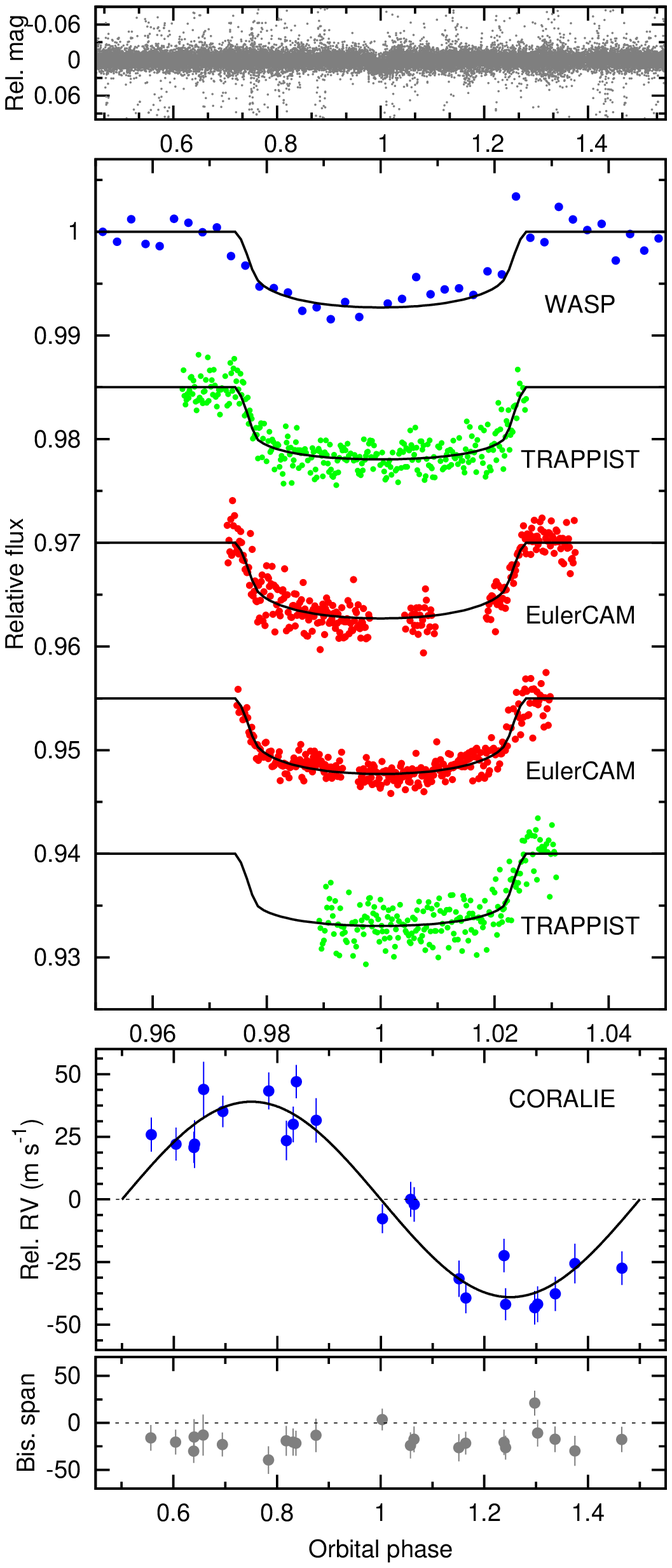}\\ [-2mm]
\caption{WASP-63b discovery data (as in Fig.~1).}
\end{figure}

\begin{table} 
\caption{System parameters for WASP-63.}
\begin{tabular}{lc}
\multicolumn{2}{l}{1SWASP\,J061720.74--381923.8}\\
\multicolumn{2}{l}{2MASS\,06172074--3819237}\\
\multicolumn{2}{l}{TYCHO-2 7612-556-1}\\
\multicolumn{2}{l}{
RA\,=\,06$^{\rm h}$17$^{\rm m}$20.74$^{\rm s}$, 
Dec\,=\,--38$^{\circ}$19$^{'}$23.8$^{''}$ (J2000)}\\
\multicolumn{2}{l}{$V$ mag = 11.2}  \\ 
\multicolumn{2}{l}{Rotational modulation\ \ \ $<$\,0.8 mmag (95\%)}\\
\multicolumn{2}{l}{pm (RA) --16.5\,$\pm$\,0.9 (Dec) --26.7\,$\pm$\,0.9 mas/yr}\\\hline
\multicolumn{2}{l}{Stellar parameters from spectroscopic analysis.\rule[-1.5mm]{0mm}{2mm}} \\ \hline 
Spectral type & G8 \\
$T_{\rm eff}$ (K)      & 5550 $\pm$ 100  \\
$\log g$      & 3.9 $\pm$ 0.1 \\
$\xi_{\rm t}$ (km\,s$^{-1}$)    & 0.9 $\pm$ 0.1 \\ 
$v\,\sin I$ (km\,s$^{-1}$)     & 2.8 $\pm$ 0.5 \\
{[Fe/H]}   &   0.08 $\pm$ 0.07 \\
{[Na/H]}   &   0.18 $\pm$ 0.06 \\ 
{[Mg/H]}   &   0.20 $\pm$ 0.05 \\
{[Si/H]}   &   0.24 $\pm$ 0.05 \\
{[Ca/H]}   &   0.18 $\pm$ 0.13 \\
{[Sc/H]}   &   0.09 $\pm$ 0.11 \\
{[Ti/H]}   &   0.12 $\pm$ 0.06 \\
{[V/H]}    &   0.16 $\pm$ 0.11 \\
{[Cr/H]}   &   0.10 $\pm$ 0.04 \\
{[Co/H]}   &   0.14 $\pm$ 0.06 \\ 
{[Ni/H]}   &   0.15 $\pm$ 0.05 \\
$\log A$(Li)  &   $<$ 0.96 $\pm$ 0.10 \\
Distance   & 330 $\pm$ 50 pc  \\ [0.5mm] \hline
\multicolumn{2}{l}{Parameters from MCMC analysis.\rule[-1.5mm]{0mm}{3mm}} \\
\hline 
$P$ (d) & 4.378090  $\pm$ 0.000006 \\
$T_{\rm c}$ (HJD) (UTC) & 2455921.6527 $\pm$ 0.0005 \\
$T_{\rm 14}$ (d) & 0.2225 $\pm$ 0.0017 \\
$T_{\rm 12}=T_{\rm 34}$ (d) & 0.017$^{+ 0.002}_{- 0.001}$\\
$\Delta F=R_{\rm P}^{2}$/R$_{*}^{2}$ & 0.00609 $\pm$ 0.00017 \\
$b$ & 0.26$^{+ 0.13}_{- 0.15}$\\
$i$ ($^\circ$)  & 87.8 $\pm$ 1.3 \\
$K_{\rm 1}$ (km s$^{-1}$) & 0.039 $\pm$ 0.003 \\
$\gamma$ (km s$^{-1}$) & --23.712 $\pm$ 0.003\\
$e$ & 0 (adopted) ($<$ 0.22 at 3$\sigma$)\\ 
$M_{\rm *}$ (M$_{\rm \odot}$) & 1.32 $\pm$ 0.05 \\
$R_{\rm *}$ (R$_{\rm \odot}$) & 1.88$^{+0.10}_{-0.06}$ \\
$\log g_{*}$ (cgs) & 4.01$^{+0.02}_{-0.04}$ \\
$\rho_{\rm *}$ ($\rho_{\rm \odot}$) & 0.198$^{+0.017}_{-0.025}$\\
$T_{\rm eff}$ (K) & 5570 $\pm$ 90\\
$M_{\rm P}$ (M$_{\rm Jup}$) & 0.38 $\pm$ 0.03 \\
$R_{\rm P}$ (R$_{\rm Jup}$) & 1.43$^{+0.10}_{-0.06}$ \\
$\log g_{\rm P}$ (cgs) & 2.62 $\pm$ 0.05 \\
$\rho_{\rm P}$ ($\rho_{\rm J}$) & 0.13 $\pm$ 0.02\\
$a$ (AU)  & 0.0574 $\pm$ 0.0007 \\
$T_{\rm P, A=0}$ (K) & 1540 $\pm$ 40 \\ [0.5mm] \hline 
\multicolumn{2}{l}{Errors are 1$\sigma$; Limb-darkening coefficients were:}\\
\multicolumn{2}{l}{(Euler $r$) a1 =    0.679, a2 = --0.433, a3 =  1.017, 
a4 = --0.494}\\
\multicolumn{2}{l}{(Trapp $Iz$) a1 = 0.766, a2 = --0.688, a3 =  
1.056, a4 = --0.479}\\ \hline
\end{tabular}
\end{table}

\newpage

\rule{0mm}{1cm}\\
\rule{0mm}{1cm}\\

\begin{figure}
\hspace*{-5mm}\includegraphics[width=10cm]{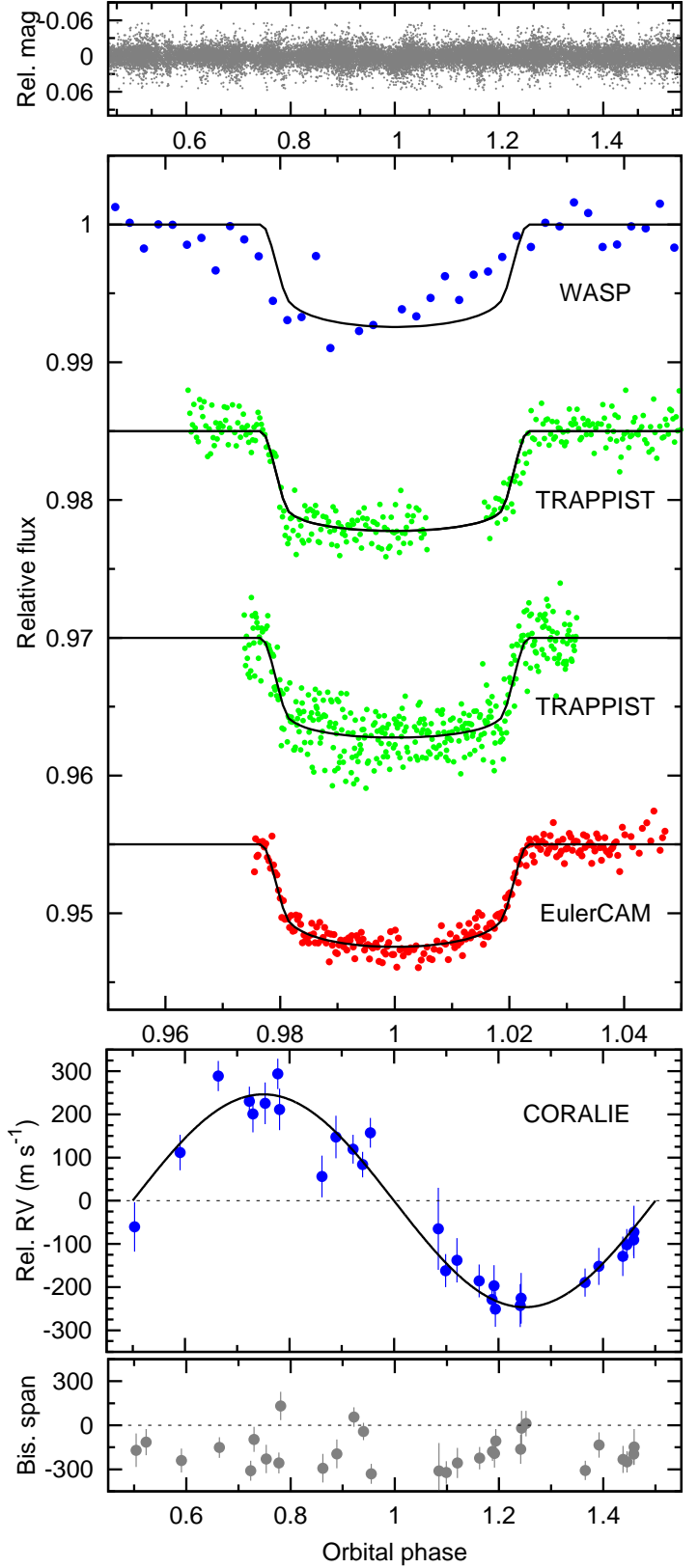}\\ [-2mm]
\caption{WASP-66b discovery data (as in Fig.~1); the two 
parts of the upper TRAPPIST curve were taken on different nights.}
\end{figure}

\begin{table} 
\caption{System parameters for WASP-66.}
\begin{tabular}{lc}
\multicolumn{2}{l}{1SWASP\,J103254.00--345923.3}\\
\multicolumn{2}{l}{2MASS\,10325399--3459234}\\
\multicolumn{2}{l}{TYCHO-2 7193-1804-1}\\
\multicolumn{2}{l}{
RA\,=\,10$^{\rm h}$32$^{\rm m}$54.00$^{\rm s}$, 
Dec\,=\,--34$^{\circ}$59$^{'}$23.3$^{''}$ (J2000)}\\
\multicolumn{2}{l}{$V$ mag = 11.6}  \\ 
\multicolumn{2}{l}{Rotational modulation\ \ \ $<$\,1 mmag}\\
\multicolumn{2}{l}{pm (RA) 11.0\,$\pm$\,0.8 (Dec) --13.1\,$\pm$\,0.8 mas/yr}\\
\hline
\multicolumn{2}{l}{Stellar parameters from spectroscopic analysis.\rule[-1.5mm]{0mm}{2mm}} \\ \hline 
Spectral type & F4 \\
$T_{\rm eff}$ (K)      & 6600 $\pm$ 150  \\
$\log g$      & 4.3 $\pm$ 0.2 \\
$\xi_{\rm t}$ (km\,s$^{-1}$)    & 2.2 $\pm$ 0.3 \\ 
$v\,\sin I$ (km\,s$^{-1}$)     & 13.4 $\pm$ 0.9 \\
{[Fe/H]}   &$-$0.31 $\pm$ 0.10 \\
{[Na/H]}   &$-$0.29 $\pm$ 0.06 \\ 
{[Mg/H]}   &$-$0.27 $\pm$ 0.10 \\
{[Si/H]}   &$-$0.19 $\pm$ 0.06 \\
{[Ca/H]}   &$-$0.19 $\pm$ 0.10 \\
{[Sc/H]}   &$-$0.17 $\pm$ 0.12 \\
{[Ti/H]}   &$-$0.16 $\pm$ 0.15 \\
{[V/H]}    &$-$0.10 $\pm$ 0.11 \\
{[Cr/H]}   &$-$0.25 $\pm$ 0.15 \\
{[Mn/H]}   &$-$0.37 $\pm$ 0.12 \\
{[Co/H]}   &$-$0.15 $\pm$ 0.08 \\ 
{[Ni/H]}   &$-$0.38 $\pm$ 0.10 \\
$\log A$(Li)[LTE] &   3.06 $\pm$ 0.11 \\
$\log A$(Li)[N-LTE]  &   2.97 $\pm$ 0.11  \\
Distance   & 380 $\pm$ 100 pc  \\ [0.5mm] \hline
\multicolumn{2}{l}{Parameters from MCMC analysis.\rule[-1.5mm]{0mm}{3mm}} \\
\hline 
$P$ (d) & 4.086052  $\pm$ 0.000007 \\
$T_{\rm c}$ (HJD) (UTC) & 2455929.09615 $\pm$ 0.00035 \\
$T_{\rm 14}$ (d) &  0.1876 $\pm$ 0.0017 \\
$T_{\rm 12}=T_{\rm 34}$ (d) & 0.018 $\pm$ 0.002\\
$\Delta F=R_{\rm P}^{2}$/R$_{*}^{2}$ & 0.00668 $\pm$ 0.00016 \\
$b$ & 0.48$^{+ 0.06}_{- 0.08}$\\
$i$ ($^\circ$)  & 85.9 $\pm$ 0.9 \\
$K_{\rm 1}$ (km s$^{-1}$) & 0.246 $\pm$ 0.011 \\
$\gamma$ (km s$^{-1}$) & --10.02458 $\pm$ 0.00013\\
$e$ & 0 (adopted) ($<$ 0.11 at 3$\sigma$)\\ 
$M_{\rm *}$ (M$_{\rm \odot}$) & 1.30 $\pm$ 0.07 \\
$R_{\rm *}$ (R$_{\rm \odot}$) & 1.75 $\pm$ 0.09  \\
$\log g_{*}$ (cgs) & 4.06 $\pm$ 0.04 \\
$\rho_{\rm *}$ ($\rho_{\rm \odot}$) & 0.242$^{+0.036}_{-0.028}$\\
$T_{\rm eff}$ (K) & 6580 $\pm$ 170\\
$M_{\rm P}$ (M$_{\rm Jup}$) & 2.32 $\pm$ 0.13 \\
$R_{\rm P}$ (R$_{\rm Jup}$) & 1.39 $\pm$ 0.09 \\
$\log g_{\rm P}$ (cgs) & 3.44 $\pm$ 0.05 \\
$\rho_{\rm P}$ ($\rho_{\rm J}$) & 0.860$^{+0.17}_{-0.13}$\\
$a$ (AU)  & 0.0546   $\pm$ 0.0009 \\
$T_{\rm P, A=0}$ (K) & 1790 $\pm$ 60 \\ [0.5mm] \hline 
\multicolumn{2}{l}{Errors are 1$\sigma$; Limb-darkening coefficients were:}\\
\multicolumn{2}{l}{(Euler $r$) a1 =    0.353, a2 = 0.759, a3 =  --0.628, 
a4 = 0.177}\\
\multicolumn{2}{l}{(Trapp $Iz$) a1 = 0.443, a2 = 0.299, a3 =  
--0.213, a4 = 0.022}\\ \hline
\end{tabular}
\end{table}

\newpage

\rule{0mm}{1cm}\\
\rule{0mm}{1cm}\\

\begin{figure}
\hspace*{-5mm}\includegraphics[width=10cm]{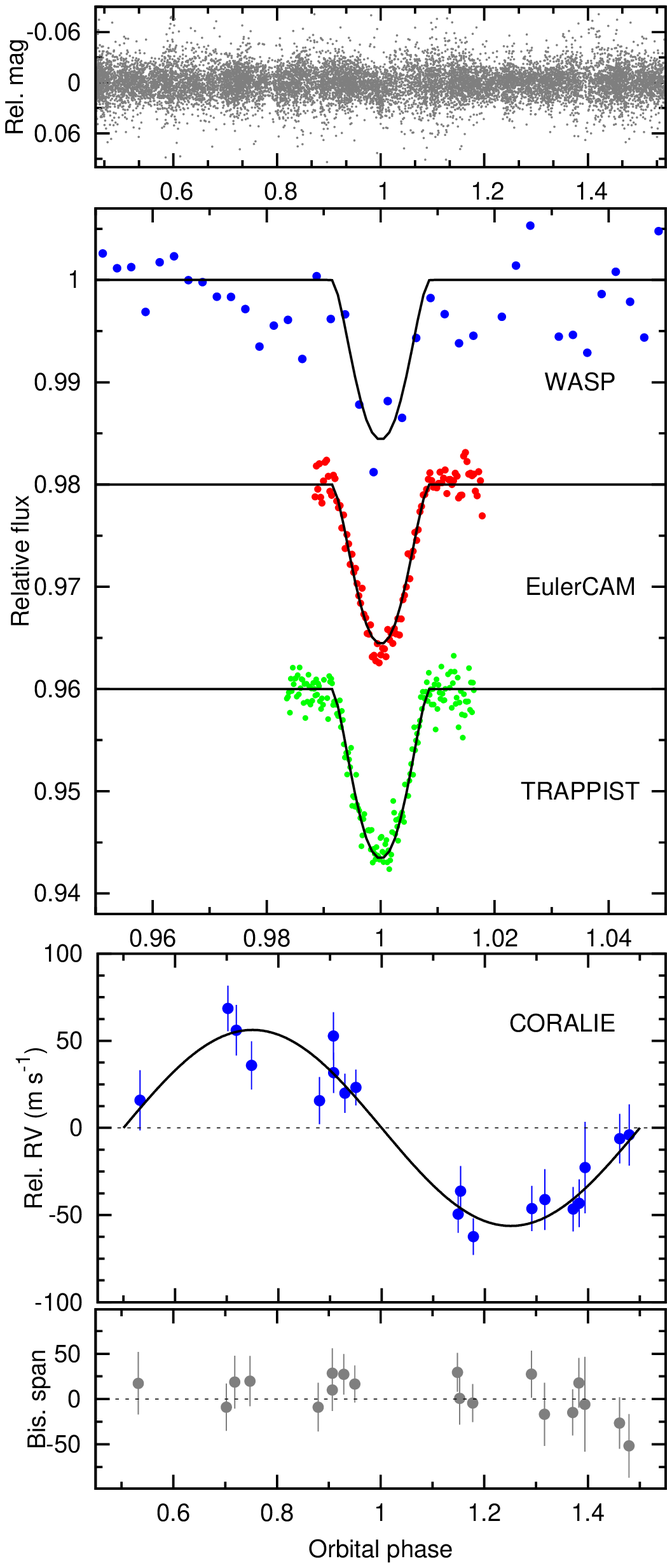}\\ [-2mm]
\caption{WASP-67b discovery data (as in Fig.~1).}
\end{figure}

\begin{table} 
\caption{System parameters for WASP-67.}
\begin{tabular}{lc}
\multicolumn{2}{l}{1SWASP\,J194258.51--195658.4}\\
\multicolumn{2}{l}{2MASS\,19425852--1956585}\\
\multicolumn{2}{l}{TYCHO-2 6307-1388-1}\\
\multicolumn{2}{l}{
RA\,=\,19$^{\rm h}$42$^{\rm m}$58.51$^{\rm s}$, 
Dec\,=\,--19$^{\circ}$56$^{'}$58.4$^{''}$ (J2000)}\\
\multicolumn{2}{l}{$V$ mag = 12.5 }  \\
\multicolumn{2}{l}{Rotational modulation\ \ \ $<$\,3 mmag}\\
\multicolumn{2}{l}{pm (RA) 0.7\,$\pm$\,1.3 (Dec) --33.8\,$\pm$\,2.4 mas/yr}\\
\hline
\multicolumn{2}{l}{Stellar parameters from spectroscopic analysis.\rule[-1.5mm]{0mm}{2mm}} \\ \hline 
Spectral type & K0V \\
$T_{\rm eff}$ (K)      & 5200 $\pm$ 100  \\
$\log g$      & 4.35 $\pm$ 0.15 \\
$\xi_{\rm t}$ (km\,s$^{-1}$)    & 0.9 $\pm$ 0.1 \\ 
$v\,\sin I$ (km\,s$^{-1}$)     & 2.1 $\pm$ 0.4 \\
{[Fe/H]}   &$-$0.07 $\pm$ 0.09 \\
{[Na/H]}   &   0.11 $\pm$ 0.08 \\ 
{[Mg/H]}   &   0.05 $\pm$ 0.04 \\
{[Si/H]}   &   0.15 $\pm$ 0.03 \\
{[Ca/H]}   &   0.02 $\pm$ 0.12 \\
{[Ti/H]}   &   0.01 $\pm$ 0.06 \\
{[V/H]}    &   0.09 $\pm$ 0.08 \\
{[Cr/H]}   &   0.06 $\pm$ 0.03 \\
{[Co/H]}   &   0.05 $\pm$ 0.04 \\  
{[Ni/H]}   &   0.00 $\pm$ 0.08 \\
$\log A$(Li)  &   $<$ 0.23 $\pm$ 0.11 \\
Distance   &   225 $\pm$ 45 pc \\ [0.5mm] \hline 
\multicolumn{2}{l}{Parameters from MCMC analysis.\rule[-1.5mm]{0mm}{3mm}} \\
\hline 
$P$ (d) & 4.61442 $\pm$ 0.00001 \\
$T_{\rm c}$ (HJD) (UTC) & 2455824.3742 $\pm$ 0.0002 \\
$T_{\rm 14}$ (d) & 0.079 $\pm$ 0.001 \\
$\Delta F=R_{\rm P}^{2}$/R$_{*}^{2}$ & 0.0181$^{+ 0.0013}_{- 0.0005}$\\
$b$ & 0.94$^{+ 0.05}_{- 0.03}$\\
$i$ ($^\circ$)  & 85.8$^{+ 0.3}_{- 0.4}$\\
$K_{\rm 1}$ (km s$^{-1}$) & 0.056 $\pm$ 0.004 \\
$\gamma$ (km s$^{-1}$)  & --0.5634 $\pm$ 0.0002 \\
$e$ & 0 (adopted) ($<$0.20 at 3$\sigma$) \\ 
$M_{\rm *}$ (M$_{\rm \odot}$) & 0.87 $\pm$ 0.04 \\
$R_{\rm *}$ (R$_{\rm \odot}$) & 0.87 $\pm$ 0.04 \\
$\log g_{*}$ (cgs) & 4.50 $\pm$ 0.03 \\
$\rho_{\rm *}$ ($\rho_{\rm \odot}$) & 1.32 $\pm$ 0.15 \\
$T_{\rm eff}$ (K) & 5240 $\pm$ 10\\ 
$M_{\rm P}$ (M$_{\rm Jup}$) & 0.42 $\pm$ 0.04 \\
$R_{\rm P}$ (R$_{\rm Jup}$) & 1.4 $^{+ 0.3}_{- 0.2}$\\
$\log g_{\rm P}$ (cgs) & 2.7$^{+ 0.1}_{- 0.2}$\\
$\rho_{\rm P}$ ($\rho_{\rm J}$) & 0.16 $\pm$ 0.08\\
$a$ (AU)  & 0.0517 $\pm$ 0.0008 \\
$T_{\rm P, A=0}$ (K) & 1040 $\pm$ 30 \\ [0.5mm] \hline
\multicolumn{2}{l}{Errors are 1$\sigma$; Limb-darkening coefficients were:}\\
\multicolumn{2}{l}{(Euler $r$) a1 =    0.671, a2 = --0.540, a3 =  1.225, 
a4 = --0.574}\\
\multicolumn{2}{l}{(Trapp $Iz$) a1 = 0.744, a2 = --0.707, a3 =  
1.134, a4 = --0.506}\\ \hline
\end{tabular} 
\end{table}

\section{WASP-47}
WASP-47 is a G9 star ($V$ = 11.9) with a possibly elevated metallicity
of [Fe/H] = 0.18 $\pm$ 0.07. There is no significant detection of
lithium in the spectra, with an equivalent width upper limit of 3m\AA,
corresponding to an abundance upper limit of $\log A$(Li) $<$ 0.81
$\pm$ 0.10. The temperature of 5400K along with the lithium abundance
imply a lower age limit of around 0.6 Gyr when compared with the Hyades
cluster (Sestito \&\ Randlich 2005). The rotation rate ($P = 15 \pm
3$~d) implied by the {\vsini} (assuming that the planet's orbit is 
aligned, and thus that the star's spin axis is perpendicular to us)
gives a gyrochronological age of $\sim
1.0^{+0.7}_{-0.4}$~Gyr using the Barnes (2007) relation.

With an orbital period of 4.16 d, a mass of 1.14 M$_{\rm Jup}$ and
a radius of 1.15 R$_{\rm Jup}$ WASP-47b is an entirely typical
hot Jupiter.  

\section{WASP-55}
WASP-55 is a G1 star ($V$ = 11.8) with a below-solar
metallicity of [Fe/H] = --0.20 $\pm$ 0.07.
The lithium abundance in WASP-55 implies an age of $\ga$ 2 Gyr
(Sestito \&\ Randlich 2005).
The rotation rate ($P = 20 \pm 7$~d) implied by the {\vsini} gives a
gyrochronological age of $\sim 3^{+5}_{-2}$~Gyr using the
Barnes (2007) relation.    

2MASS images of WASP-55
show a star approximately 2$^{\prime\prime}$ away and about
5 magnitudes fainter.  This is sufficiently faint that it is
unlikely to be affecting our results significantly. 

WASP-55b is moderately inflated, with a mass of 0.57 M\jup\ and
a radius of 1.30 R\jup, though this is in line with many known
hot Jupiters.

\section{WASP-61}
WASP-61 is an F7 star ($V$ = 12.5) with metallicity near solar (the
poor quality of our spectrum prevents more detailed analysis than the
[Fe/H] = --0.10 $\pm$ 0.12 reported in Table~4).  There is no
significant detection of lithium in the spectra, corresponding to an
abundance upper limit of $\log A$(Li) $<$ 1.1 $\pm$ 0.1, which implies
an age of several Gyr (Sestito \&\ Randlich 2005). The rotation
rate ($P = 6.3 \pm 0.9$~d) implied by the {\vsini} gives a
gyrochronological age of $\sim 0.7^{+1.2}_{-0.4}$~Gyr using the
Barnes (2007) relation.  

WASP-61b has a high mass of $M$ = 2.1 M\jup\ and the highest density of the 
planets reported here, at $\rho$ = 1.1 $\rho$\jup.

\section{WASP-62}
WASP-62 is an F7 star ($V$ = 10.3) with a solar metallicity.  For a
star of this temperature (6230 $\pm$ 80 K) the presence of relatively
strong lithium absorption in the spectrum does not provide a strong
age constraint; this level of depletion is found in clusters as young
as $\sim$0.5~Gyr (Sestito \&\ Randlich 2005). The rotation rate ($P =
6.3 \pm 0.8$~d) implied by the {\vsini} gives a gyrochronological age
of $\sim 0.7^{+0.4}_{-0.3}$~Gy using the Barnes (2007)
relation. There are no emission peaks evident in the Ca~{\sc ii} H+K
lines. 

The EulerCAM transit lightcurve is badly affected by weather.
Our MCMC analysis balances $\chi^{2}$ across the different 
datasets, so inflates the error bars of this lightcurve.  We
also ran the analysis omitting this curve, which led to results
that were the same within the errors.

\section{WASP-63}
WASP-63 is a G8 star ($V$ = 11.2) with solar metallicity.  There is no
significant detection of lithium in the spectra, with an equivalent
width upper limit of 11m\AA, corresponding to an abundance upper limit
of $\log A$(Li) $<$ 0.96 $\pm$ 0.10. This implies an age of at least 
several Gyr (Sestito \&\ Randlich 2005).  The rotation rate ($P = 37
\pm 9$~d) implied by the {\vsini} gives a gyrochronological age of
$\sim 6^{+5}_{-3}$~Gyr using the Barnes (2007)
relation.  There are no emission peaks evident in the Ca~{\sc ii} H+K
lines.  

The stellar radius is  inflated for a G8 star and 
indicates that WASP-63 has evolved off the main sequence (see 
Fig.~8), with an age of $\sim$ 8 Gyr.    

The planet 
WASP-63b is the least massive of those
reported here, at 0.38 M\jup, and also has the lowest density
($\rho$ = 0.13 $\rho$\jup).  This indicates that mechanisms causing
inflated planet radii need to be able to operate late on in the 
evolution of a planetary system.  

\section{WASP-66}
WASP-66 is an F4 star ($V$ = 11.6) with a below-solar metallicity of
[Fe/H] = --0.31 $\pm$ 0.10.  With $T_{\rm eff}$ = 6600\,$\pm$\,150 K
WASP-66 is relatively hot among known hot-Jupiter hosts.  
The presence of strong lithium absorption
in the spectrum suggests that WASP-66 is $\la$2~Gyr old (Sestito \&\
Randlich 2005). The rotation rate ($P = 4.9 \pm 1.3$~d) implied by the
{\vsini} gives little age constraint, $3.3^{+10}_{-2.7}$~Gyr, from the
Barnes (2007) relation.

With a mass of 2.3 M\jup\ WASP-66b is the most massive of the
planets reported here.  

\section{WASP-67}
WASP-67 is an K0V star ($V$ = 12.5) with a solar metallicity. 
There is no significant detection of lithium in the spectra, with an
equivalent width upper limit of 5m\AA, corresponding to an abundance
upper limit of $\log A$(Li) $<$ 0.23 $\pm$ 0.11. This implies an age
of at least $\sim$0.5~Gyr (Sestito \&\ Randlich 2005).  The rotation rate
($P = 25 \pm 7$~d) implied by the {\vsini} gives a
gyrochronological age of $2.0^{+1.6}_{-1.0}$~Gyr using the
Barnes (2007) relation.  There are no emission peaks
evident in the Ca~{\sc ii} H+K lines.  

WASP-67b has a mass of 0.42 M\jup\ and a radius of 1.4 M\jup, making
it inflated ($\rho$ = 0.16 $\rho$\jup).  It also has a high impact
factor of $b = 0.94$, making the transit curve V-shaped.  If the
criterion $X = b + R_{\rm P}/R_{*} > 1$ is satisfied then the transit
is grazing, with part of the planet not transiting the stellar face
(see, e.g., Smalley \etal\ 2011). For WASP-67b this value is
1.07$^{+0.05}_{-0.03}$, and, further, out of 375\,000 MCMC steps only 17
had $X < 1$.  This implies a $>$3$\sigma$ probability that the transit is
grazing, making WASP-67b the first hot Jupiter known to have a 
grazing transit, following WASP-34 and HAT-P-27/WASP-40 that are
possibly grazing (Smalley \etal\ 2011; Anderson \etal\ 2011).

\begin{figure}
\hspace*{-5mm}\includegraphics[width=9cm]{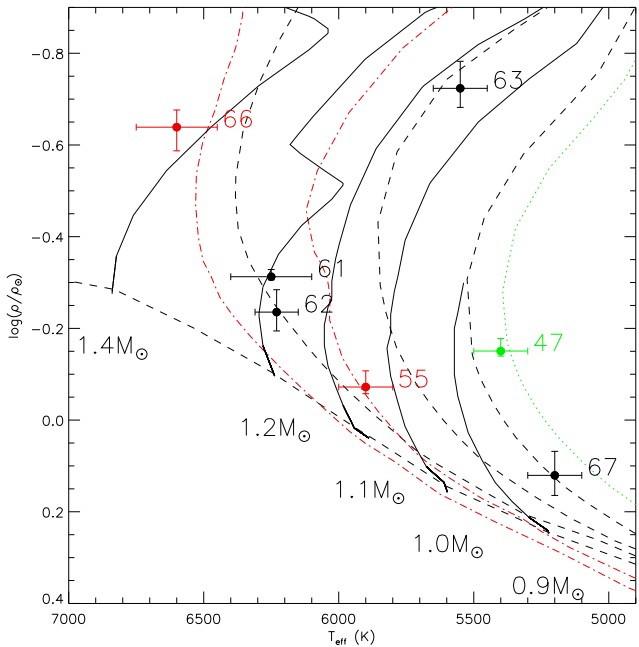}
\caption{Evolutionary tracks on a modified H--R diagram
($\rho_{*}^{-1/3}$ versus $T_{\rm eff}$).
The black lines are for solar metallicity, [Fe/H] = 0,
showing (solid lines) mass tracks with the labelled masses,
and (dashed lines) age tracks for $\log$(age) = 7.8, 9.4, 9.9 
\&\ 10.2 yrs. The four host stars with near-solar metallicities
are shown in black.   
The green-labelled WASP-47 has above-solar metallicity
of [Fe/H] = +0.18, and the green dotted line is the
age track for [Fe/H] = +0.18 and $\log$(age) = 10.2 yrs. 
The red dot-dashed lines are for a below-solar metallicity
of [Fe/H] = --0.2, being at ages $\log$(age) = 9.4 \&\ 9.8 yrs.
The two host stars with below-solar metallicities are
labelled in red.  The data are from Girardi \etal\ (2000).}
\end{figure}

\section{Discussion}
It has often been noted that the hot-Jupiter population
shows an apparent ``pile up'' at orbital periods of $P$ = 3--4 d.
We can use the increasing numbers of hot Jupiters, primarily
from the ground-based transit surveys, to investigate this.

\begin{figure}
\includegraphics[width=9cm]{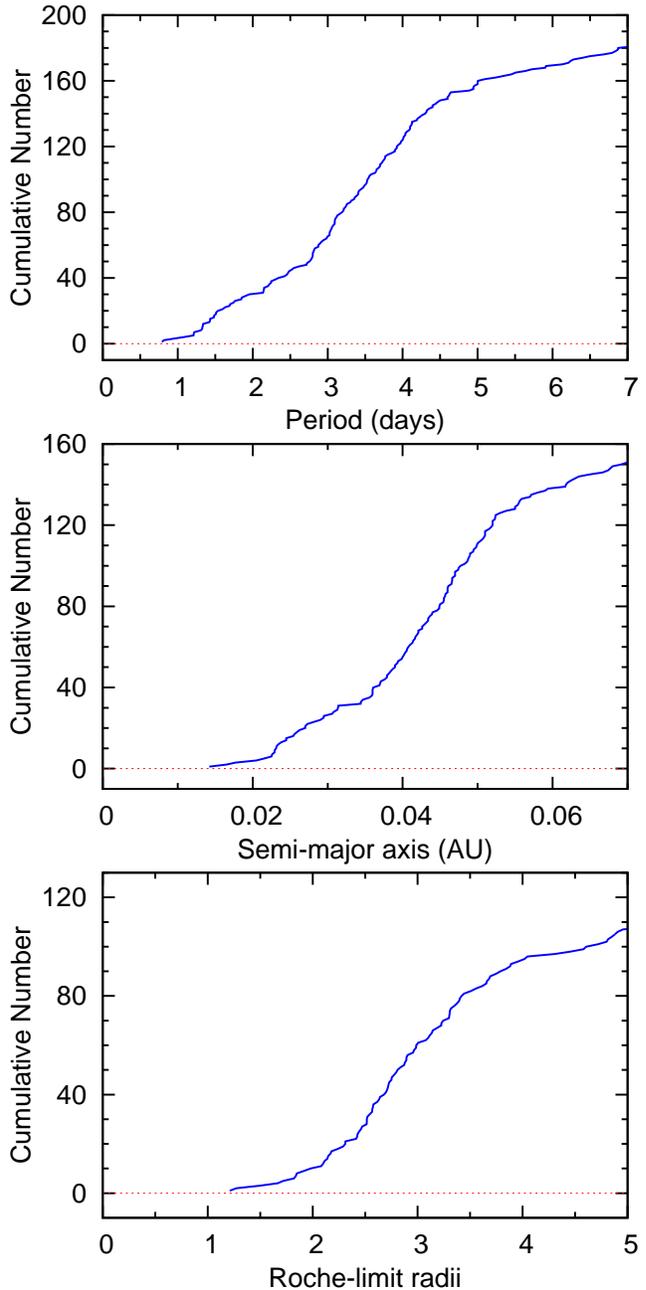}
\caption{The cumulative distributions of (top) orbital period,
(middle) semi-major axis, and (bottom) separations in units
of Roche-limit separation for a sample of hot Jupiters (see text).  
The definition of Roche-limit separation is taken from Ford \&\ Rasio (2006).} 
\end{figure}

\begin{figure}
\includegraphics[width=9cm]{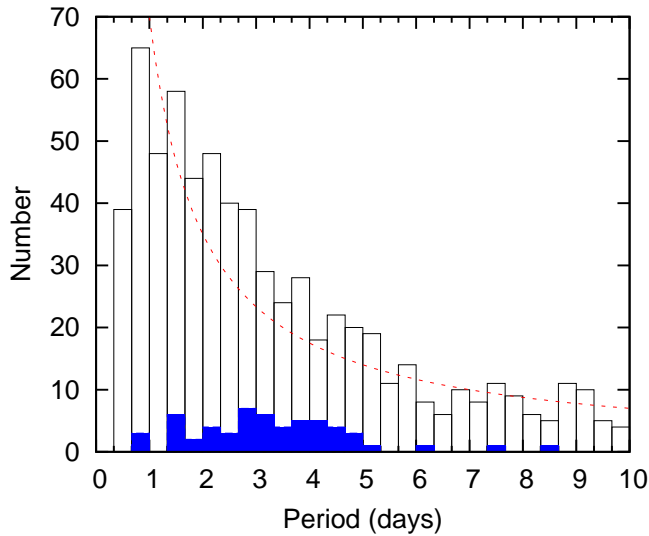}
\caption{The open histogram shows the period distribution
of all WASP-South candidates that have been rejected or
confirmed as planets.  The solid histogram shows the planets.
The dotted line is a simple $P^{-1}$ function, scaling
as the number of transits observed in a WASP-like survey.}
\end{figure}

As a first step we take the sample of confirmed planets
compiled by Schneider \etal\ (2011), as of March 2012, limiting this 
to periods less than 8 days and planetary masses of 0.1--12 M$_{\rm Jup}$ 
(the super-Earths may well be a different population dynamically).
We show (Fig.~9) the cumulative distributions against
orbital period, semi-major axis and Roche-limit separation.  These
confirm that we see more planets at periods $P$ of
$\approx$\,3--5 days, with fewer at shorter and longer periods. 
However, this compilation comes from many different surveys,
each of which will have different selection effects, and so
needs to be interpreted with caution.  

We thus create a second sample of planets discovered by
the transit surveys ($P < 8$\,d, $M$ = 0.1--12 M$_{\rm Jup}$), based
on the Schneider \etal\ compilation but with unpublished WASP
planets added up to WASP-84b.  This sample of 163 planets 
is dominated by WASP (81 planets), HAT (34), Kepler (18) and CoRoT (14). 

The inclination range that produces a transit scales with
semi-major axis as $\cos^{-1}(R_{*}/a)$. To compare this with
the distribution of orbital periods, which are securely known, 
we can translate this to $P$ by assuming a star of solar
mass and radius.  Further, the biggest factor affecting
discovery probability in a WASP-like survey 
is the number of transits recorded, which will scale as
$P^{-1}$.  In Fig.~10 we show the distribution of rejected
WASP-South candidates, which indicates that a $P^{-1}$ function,
though imperfect, is a rough approximation. 

We caution that this is only a very preliminary account of
relevant selection effects, which will be different for each
of the above surveys.  For example, the number of transits 
required is likely to saturate above some number (this number 
depending on the survey and the amount of data),
and WASP-like surveys at only one longitude will also suffer
from sampling effects at integer-day periods.   

Nevertheless, we can multiply together the transit probability
and the $P^{-1}$ function to produce the detection-probability 
curve shown in Fig.~11, and we can use this to produce
a ``corrected'' planet distribution curve.

One can interpret this curve as showing four regions with different
slopes, the slopes having relative ratios (for corrected number of
planets versus period interval) of 1\,:\,10\,:\,40\,:\,12.  
The need for different
slopes in the different regions is significant on a K--S test at
$>$95\%\ probabilities.  We caution, though, that we regard this as an
indicative description of the distribution, rather than a unique one,
and the relative slopes are of course dependent on the uncertain
selection effects.

The four regions of the hot-Jupiter 
period distribution are:

(1) $P$ = 0.8--1.2 d, containing only 
4 planets (WASP-19b, WASP-43b, WASP-18b \&\ WASP-12b;
Hebb \etal\ 2010; Hellier \etal\ 2011b; Hellier \etal\ 2009; Hebb \etal\ 2009),
despite the probability of detection being greatest.   
These planets are thought to be tidally decaying on 
relatively short timescales, and so are rare, found
only by the surveys sampling the most stars 
(see the discussion in Hellier \etal\ 2011b;  
note that WASP observes from one longitude with greater sky coverage 
than HATnet, whereas HATnet covers less sky but from
several longitudes). 

There are no known hot Jupiters with a period
below the $P$ = 0.79-d of WASP-19b. Despite the
fact that the probability of detection of such planets
in WASP data is at its highest (see Fig.~11 curve), the 
number of good candidates declines (Fig.~10), and, further, we have 
followed up over 40 such candidates without success 
(compared to an overall success rate of 1-in-12). 
Thus hot-Jupiter planets below $P$ = 0.79-d must be
very rare (there are several super-Earths with 
such periods, though their tidal-decay rate will of course
be much lower). 

\begin{figure}
\includegraphics[width=9cm]{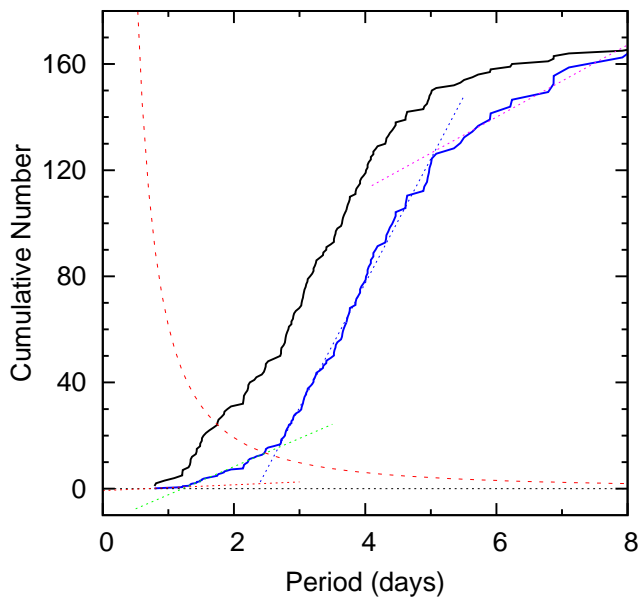}
\caption{The solid-black line is the cumulative distribution
of orbital periods for transiting hot Jupiters. The solid-blue
line is the same but corrected for the probability of
detection (which is the dotted curve). The dotted straight
lines are a suggested parametrisation of the distribution
(see text). }
\end{figure}

(2) $P$ = 1.2--2.7 d. The abundance is an order
of magnitude greater than in the 0.8--1.2 d range,
but still a factor $\sim$\,4 lower than in
the range 2.7--5-d.

(3) $P$ = 2.7--5-d.  Our analysis confirms the 
existence of a pile-up of hot Jupiters, and
suggests that it has a relatively well defined lower edge 
at $P$ = 2.7-d.   

Ford \&\ Rasio (2006) argued that a hot-Jupiter population 
resulting from circularisation of highly eccentric orbits will 
have an inner edge at 2 Roche-limit  radii (2$a_{R}$ where
$a_{R} = \frac{R_{P}}{0.462}\left(\frac{M_{P}}{M_{*}}\right)^{1/3}$).

For a planet of Jupiter mass and radius
around a star of solar mass and radius, 2\,$a_{R}$ 
corresponds to $P$ = 1.2-d, and so would explain our
finding of a break at that period.  The few systems 
inside that limit are presumably spiralling inward 
relatively rapidly under 
tidal decay (e.g.\  Matsumura, Peale \&\ Rasio 2010).   

Further, for an inflated planet with a radius
of $R \approx 1.8$\,R\jup\ (as seen in the highly
inflated planets WASP-12b, WASP-17b, HAT-P-32b
\&\ HAT-P-33b; Hebb \etal\ 2009; Anderson \etal\ 2010; 
Hartman \etal\ 2011) 2\,$a_{R}$ equates to 
$P$ = 2.7 d (again assuming a 1-M\jup\ planet orbiting 
a sun-like star).  Hence, if hot Jupiters arrive in the 
pile-up with a range of radius inflations, the Ford \&\ 
Rasio argument would produce a cut-off ranging from 
$\sim$\,1.2 to 2.7 d, which might explain our finding
of breaks at both those values. 

(4) $P$ = \sqiggt 5-d.  The upper edge of the 
pile-up appears to be near 5-d, although we 
caution that in WASP-like surveys the detection probability 
(and hence number of planets) decreases, and the selection
effects get worse, as the period increases beyond $P$\,$\sim$\,5 d.
For this reason we don't further interpret the longer-period  
range. 

The period distribution of hot Jupiters is likely
to result from several physical mechanisms.  These
include disk migration and possible `stopping mechanisms' 
(e.g.\ Matsumura, Pudritz \&\ Thommes 2007), 
third-body interactions, such as the Kozai mechanism,
that can move planets onto highly eccentric orbits that 
are then tidally captured and circularise
at short periods (e.g.\ Guillochon \etal\ 2011;
Noaz \etal\ 2011), and
orbital decay and in-spiral caused by tidal interactions 
with the host star (e.g.\ Matsumura \etal\ 2010). 

Study of the angle between the planetary orbit and 
the stellar rotation axis indicates that many current
orbits are likely to result from the Kozai mechanism 
(e.g.\ Triaud \etal\ 2010), but it is probable that the hot Jupiters 
are a composite population with differing 
past histories.    Thus, to further investigate the 
pile-up, we need to accumulate statistics
to look for differences in, for example, the orbital
eccentricities and the spin--orbit angle between the different
period ranges that we have outlined.

\section*{Acknowledgements}
WASP-South is hosted by the South African
Astronomical Observatory and
we are grateful for their ongoing support and assistance. 
Funding for WASP comes from consortium universities
and from the UK's Science and Technology Facilities Council.
TRAPPIST is funded by the Belgian Fund for Scientific  
Research (Fond National de la Recherche Scientifique, FNRS) under the  
grant FRFC 2.5.594.09.F, with the participation of the Swiss National  
Science Fundation (SNF).  M. Gillon and E. Jehin are FNRS Research  
Associates.

\newpage
\appendix

\section{Online only} 

\begin{table}
\caption{CORALIE radial velocities.\protect\rule[-1.5mm]{0mm}{2mm}} 
\begin{tabular}{cccr} 
\hline 
BJD\,--\,2400\,000 & RV & $\sigma$$_{\rm RV}$ & Bisector \\
(UTC)  & (km s$^{-1}$) & (km s$^{-1}$) & (km s$^{-1}$)\\ [0.5mm] \hline
\multicolumn{4}{l}{{\bf WASP-47:}}\\  
55328.9147 & $-$27.199 & 0.010 & $-$0.050\\
55384.9303 & $-$26.943 & 0.015 & $-$0.056\\
55389.7034 & $-$27.008 & 0.010 & $-$0.074\\
55390.7133 & $-$27.186 & 0.011 & $-$0.041\\
55391.8362 & $-$27.142 & 0.008 & $-$0.044\\
55392.6807 & $-$26.980 & 0.009 & $-$0.019\\
55396.7987 & $-$26.954 & 0.010 & $-$0.041\\
55403.6889 & $-$27.201 & 0.009 & $-$0.080\\
55408.6869 & $-$27.107 & 0.015 & $-$0.064\\
55409.6504 & $-$26.959 & 0.009 & $-$0.032\\
55450.5481 & $-$27.054 & 0.012 & $-$0.077\\
55454.7270 & $-$27.055 & 0.009 & $-$0.029\\
55782.7653 & $-$27.121 & 0.011 & $-$0.061\\
55795.7802 & $-$27.003 & 0.011 & $-$0.040\\
55809.5958 & $-$26.937 & 0.011 & $-$0.065\\
55885.5702 & $-$27.170 & 0.016 & $-$0.049\\
55886.5550 & $-$27.153 & 0.014 & $-$0.026\\
55887.6093 & $-$26.937 & 0.013 & $-$0.007\\
55888.5579 & $-$26.941 & 0.011 & $-$0.029\\ [0.5mm]
\multicolumn{4}{l}{{\bf WASP-55:}}\\ 
55591.7843 & $-$4.414 & 0.020 & ~0.036\\
55593.8655 & $-$4.238 & 0.012 & $-$0.003\\
55595.8370 & $-$4.375 & 0.014 & $-$0.005\\
55596.7857 & $-$4.352 & 0.013 & ~0.019\\
55598.7838 & $-$4.287 & 0.012 & $-$0.021\\
55599.8011 & $-$4.360 & 0.019 & $-$0.039\\
55602.8371 & $-$4.243 & 0.013 & $-$0.022\\
55604.8196 & $-$4.377 & 0.013 & $-$0.000\\
55605.7954 & $-$4.368 & 0.013 & $-$0.028\\
55607.7461 & $-$4.275 & 0.014 & $-$0.047\\
55623.7577 & $-$4.356 & 0.017 & ~0.006\\
55624.7801 & $-$4.262 & 0.014 & $-$0.020\\
55625.6823 & $-$4.264 & 0.014 & $-$0.009\\
55629.8416 & $-$4.271 & 0.013 & $-$0.033\\
55635.7567 & $-$4.367 & 0.021 & ~0.050\\
55637.7936 & $-$4.308 & 0.024 & $-$0.097\\
55665.5870 & $-$4.240 & 0.014 & ~0.006\\
55675.8082 & $-$4.384 & 0.016 & $-$0.040\\
55679.7306 & $-$4.333 & 0.014 & $-$0.013\\
55764.5358 & $-$4.290 & 0.015 & 0.044\\ [0.5mm]
\multicolumn{4}{l}{{\bf WASP-61:}}\\ 
55570.7149 & 18.857 & 0.041 & ~0.099\\
55615.6221 & 19.218 & 0.039 & ~0.009\\
55629.5093 & 18.790 & 0.046 & $-$0.054\\
55649.5172 & 18.985 & 0.045 & 0.080\\
55650.5243 & 19.168 & 0.048 & $-$0.020\\
55802.8748 & 18.721 & 0.056 & $-$0.120\\
55806.8968 & 18.758 & 0.039 & ~0.038\\
55809.8327 & 18.748 & 0.049 & ~0.132\\
55810.8849 & 18.772 & 0.045 & ~0.017\\
55811.8571 & 19.108 & 0.050 & ~0.140\\
55813.8869 & 18.810 & 0.057 & $-$0.004\\
55814.9060 & 18.891 & 0.050 & ~0.106\\
55815.8979 & 19.172 & 0.049 & ~0.148\\
55839.8379 & 19.172 & 0.046 & $-$0.024\\
55868.8371 & 18.923 & 0.038 & $-$0.015\\ [0.5mm]
\hline
\multicolumn{4}{l}{Bisector errors are twice RV errors} 
\end{tabular} 
\end{table} 

\begin{table}
\addtocounter{table}{-1}
\caption{continued\protect\rule[-1.5mm]{0mm}{2mm}} 
\begin{tabular}{cccr} 
\hline 
BJD\,--\,2400\,000 & RV & $\sigma$$_{\rm RV}$ & Bisector \\
(UTC)  & (km s$^{-1}$) & (km s$^{-1}$) & (km s$^{-1}$)\\ [0.5mm] \hline
\multicolumn{4}{l}{{\bf WASP-62:}}\\  
55651.4898 & 15.039 & 0.009 & ~0.043	 \\
55675.4985 & 14.921 & 0.010 & ~0.022	\\
55676.4710 & 14.963 & 0.009 & --0.006\\
55677.4992 & 15.023 & 0.009 & --0.005\\
55678.4947 & 14.985 & 0.010 & ~0.030	\\
55681.4724 & 15.017 & 0.010 & ~0.035\\
55682.4765 & 15.032 & 0.010 & --0.038\\
55683.4807 & 14.963 & 0.010 & ~0.020	\\
55684.4733 & 14.899 & 0.009 & ~0.054	\\
55685.4723 & 14.969 & 0.011 & ~0.066\\
55692.4975 & 14.953 & 0.013 & --0.013\\
55693.4947 & 14.928 & 0.010 & ~0.005	\\
55809.8571 & 15.027 & 0.012 & ~0.033\\
55810.9091 & 14.979 & 0.011 & ~0.042\\
55811.8813 & 14.922 & 0.015 & ~0.026\\
55815.8730 & 14.924 & 0.010 & ~0.061\\
55836.8736 & 14.998 & 0.010 & ~0.050\\
55839.8625 & 14.957 & 0.010 & --0.003\\
55840.8861 & 15.017 & 0.010 & ~0.024	\\
55850.8434 & 14.981 & 0.011 & ~0.065\\
55864.8453 & 14.910 & 0.009 & ~0.012\\
55881.8528 & 14.979 & 0.010 & --0.041\\
55888.6963 & 14.991 & 0.011 & ~0.012	\\
55889.6691 & 15.056 & 0.010 & ~0.059\\
56021.5148 & 15.039 & 0.012 & ~0.048  \\ [0.5mm]
\multicolumn{4}{l}{{\bf WASP-63:}}\\    
55598.7288 & --23.771 & 0.006 & --0.026 \\ 
55649.4954 & --23.679 & 0.007 & --0.022 \\ 
55651.5371 & --23.768 & 0.007 & --0.011 \\ 
55670.5213 & --23.705 & 0.006 & --0.030 \\ 
55675.5315 & --23.682 & 0.007 & --0.040 \\ 
55676.4936 & --23.733 & 0.006 & ~0.003 \\ 
55677.5218 & --23.747 & 0.007 & --0.021  \\ 
55678.5172 & --23.752 & 0.007 & --0.018  \\ 
55679.5209 & --23.690 & 0.006 & --0.023 \\ 
55681.5178 & --23.756 & 0.007 & --0.026 \\ 
55682.4989 & --23.750 & 0.008 & --0.030  \\ 
55683.5030 & --23.703 & 0.007 & --0.020 \\ 
55684.4956 & --23.695 & 0.007 & --0.020 \\ 
55685.5179 & --23.727 & 0.007 & --0.017 \\ 
55695.4660 & --23.762 & 0.007 & --0.017 \\ 
55802.8982 & --23.687 & 0.009 & --0.013 \\ 
55805.8813 & --23.692 & 0.007 & --0.016  \\ 
55836.8931 & --23.694 & 0.010 & --0.015 \\ 
55858.8607 & --23.671 & 0.011 & --0.013  \\ 
55869.8341 & --23.754 & 0.006 & --0.022 \\ 
55895.6363 & --23.713 & 0.007 & --0.024  \\ 
55966.7350 & --23.752 & 0.007 & ~0.021 	 \\ 
56021.5498 & --23.683 & 0.008 & --0.019 \\ [0.5mm] 
\hline
\multicolumn{4}{l}{Bisector errors are twice RV errors} 
\end{tabular} 
\end{table} 

\begin{table}
\addtocounter{table}{-1}
\caption{continued\protect\rule[-1.5mm]{0mm}{2mm}} 
\begin{tabular}{cccr} 
\hline 
BJD\,--\,2400\,000 & RV & $\sigma$$_{\rm RV}$ & Bisector \\
(UTC)  & (km s$^{-1}$) & (km s$^{-1}$) & (km s$^{-1}$)\\ [0.5mm] \hline
\multicolumn{4}{l}{{\bf WASP-66:}}\\  
55572.7155 & ~--9.813 & 0.048 & ~0.131 \\ 
55623.6364 & --10.250 & 0.059 & --0.019 \\ 
55627.7156 & --10.267 & 0.050 & --0.163 \\ 
55628.7858 & --10.085 & 0.057 & --0.170 \\ 
55629.7107 & ~--9.824 & 0.043 & --0.096 \\ 
55632.6904 & --10.097 & 0.061 & --0.148 \\ 
55635.6818 & --10.222 & 0.048 & --0.192 \\ 
55643.5642 & --10.162 & 0.051 & --0.257 \\ 
55644.6746 & --10.177 & 0.043 & --0.135 \\ 
55646.7038 & ~--9.877 & 0.049 & --0.195 \\ 
55647.5623 & --10.186 & 0.038 & --0.322 \\ 
55648.6521 & --10.214 & 0.033 & --0.307 \\ 
55649.5716 & ~--9.913 & 0.041 & --0.240 \\ 
55650.6813 & ~--9.968 & 0.048 & ~0.292 \\ 
55651.5910 & --10.089 & 0.095 & --0.311 \\ 
55674.6304 & ~--9.794 & 0.034 & --0.310 \\ 
55675.5760 & ~--9.867 & 0.034 & --0.331 \\ 
55676.5274 & --10.254 & 0.033 & --0.179 \\ 
55677.6348 & --10.115 & 0.037 & --0.196 \\ 
55679.6014 & ~--9.941 & 0.030 & --0.043 \\ 
55680.6394 & --10.276 & 0.041 & --0.108 \\ 
55681.6674 & --10.126 & 0.036 & --0.250 \\ 
55682.5584 & ~--9.736 & 0.035 & --0.152 \\ 
55683.6124 & ~--9.905 & 0.033 & ~0.055 \\ 
55684.5982 & --10.210 & 0.038 & --0.223 \\ 
55707.5410 & ~--9.731 & 0.035 & --0.257 \\ 
55711.5280 & ~--9.799 & 0.048 & --0.229 \\ 
55722.4965 & --10.153 & 0.046 & --0.232 \\ 
55958.7274 & --10.388 & 0.044 & ~0.012 \\ 
56000.6947 & ~--9.983 & 0.045 & --0.115 \\ [0.5mm] 
\multicolumn{4}{l}{{\bf WASP-67:}}\\  
55765.7296 & $-$0.610 & 0.013 & ~0.028\\
55767.7027 & $-$0.507 & 0.015 & ~0.019\\
55768.6736 & $-$0.544 & 0.011 & ~0.027\\
55769.6850 & $-$0.613 & 0.011 & ~0.030\\
55770.7119 & $-$0.610 & 0.013 & $-$0.015\\
55777.7996 & $-$0.511 & 0.014 & ~0.029\\
55793.5324 & $-$0.605 & 0.017 & $-$0.017\\
55794.5285 & $-$0.548 & 0.017 & ~0.017\\
55795.5247 & $-$0.528 & 0.014 & ~0.020\\
55798.5058 & $-$0.586 & 0.026 & $-$0.006\\
55803.5145 & $-$0.567 & 0.018 & $-$0.052\\
55804.5425 & $-$0.495 & 0.013 & $-$0.009\\
55805.4867 & $-$0.532 & 0.012 & ~0.010\\
55806.6210 & $-$0.600 & 0.015 & ~0.001\\
55807.6815 & $-$0.607 & 0.014 & ~0.018\\
55819.5305 & $-$0.540 & 0.010 & ~0.017\\
55820.5790 & $-$0.626 & 0.010 & $-$0.004\\
55826.5015 & $-$0.570 & 0.014 & $-$0.026\\
55851.5050 & $-$0.548 & 0.013 & $-$0.009\\
\hline
\multicolumn{4}{l}{Bisector errors are twice RV errors} 
\end{tabular} 
\end{table}


\begin{thebibliography}{}
\bibitem[Anderson \etal\ (2010)]{w17}Anderson D. R. \etal, 2010, ApJ, 709, 159
\bibitem[Anderson \etal\ (2011)]{w40}Anderson D. R. \etal, 2011, PASP, 123, 555
\bibitem[Anderson \etal\ (2012)]{w44}Anderson D. R. \etal, 2012, MNRAS, in press (arXiv1105.3179)
\bibitem[Bakos et al.(2004)]{hat}Bakos G. \`{A}., Noyes R. W., Kov\'acs G., Stanek K. Z., Sasselov, D. D., Domsa, I., PASP, 116, 266
\bibitem[Barnes (2007)]{2007ApJ...669.1167B} Barnes, S.A. 2007, ApJ, 669, 1167
\bibitem[Batalha et al. (2012)]{batalha}Batalha N. M. \etal, 2012, ApJS submitted (arXiv1202.5852)
\bibitem[Bruntt et al. (2010)]{2010MNRAS.405.1907B} Bruntt, H. et al., 2010, MNRAS, 405, 1907
\bibitem[Claret (2000)]{claret}Claret, A., 2000, A\&A, 363, 1081
\bibitem[Collier Cameron et al.(2007a)]{wasp1}Collier Cameron, A., et al., 2007a, MNRAS, 375, 951 
\bibitem[Collier Cameron et al.(2007b)]{mcmc}Collier Cameron, A., et al., 2007b, MNRAS, 380, 1230 
\bibitem[Enoch \etal\ (2010)]{enoch}Enoch B., Collier Cameron A., Parley N.~R., Hebb L., 2010, A\&A, 516, A33
\bibitem[Ford \& Rasio (2006)]{fordrasio}Ford, E. B. \&\ Rasio, F. A., 2006, 
ApJ, 638, L45
\bibitem[Gillon et al.\ (2007)]{spectral}Gillon, M., et al., 2009, A\&A, 496, 259
\bibitem{giradi}Girardi, L., Bressan, A., Bertelli, G., Chiosi, C. 2000, A\&AS,
 141, 371
\bibitem[Guillochon et al.\ (2011)]{guill}Guillochon, J., Ramirez-Ruiz, E., Lin, D. N. C., 2011,  ApJ, 732, 74 
\bibitem[Hebb et al.\ (2009)]{hebb10}Hebb, L. et al., 2009, ApJ, 693, 1920
\bibitem[Hebb et al.\ (2010)]{hebb09}Hebb, L. et al., 2010, ApJ, 708, 224
\bibitem[Hellier et al.\ (2009)]{w18}Hellier, C. et al., 2009, Nature, 460, 1098
\bibitem[Hellier et al.\ (2011a)]{hohp}Hellier, C. et al., 2011a, in 
``Detection and dynamics of transiting exoplanets", eds F. Bouchy, R. D\'iaz,
 C. Moutou, EPJ Web of Conferences, Volume 11, id.01004
\bibitem[Hellier et al.\ (2011b)]{w43}Hellier, C. et al., 2011b, A\&A, 535, L7
\bibitem[Jehin et al.\ (2011)]{trappist}Jehin, E. et al., 2011, Messenger, 145, 2
\bibitem{mats}Matsumura, S., Pudritz, R.E.. Thommes, E.W. 2007, ApJ, 660, 1609
\bibitem[Matsumura et al.\ (2010)]{matsumura}Matsumura, S., Peale, S. J., Rasio, F. A., 2010, ApJ, 725, 1995
\bibitem{maxted}Maxted, P.F.L. \etal\ 2011, PASP, 123, 547 
\bibitem{naoz}Naoz, S., Farr, W. M., Lithwick, Y., Rasio, F. A., Teyssandier, J., 2011, Nature, 473, 187
\bibitem{navarro}Navarro J. F., Abadi M. G.. Venn, K. A., Freeman K. C., Anguiano B., 2011, MNRAS, 412, 1203
\bibitem[Magain (1984)]{magain}Magain, P., 1984, A\&A, 134, 189
\bibitem[Pollacco et al.\ (2006)]{sw}Pollacco, D., et al., 2006, PASP, 118, 1407 
\bibitem[Pollacco et al.\ (2007)]{wasp3}Pollacco, D., et al., 2008, MNRAS, 385, 1576
\bibitem[Schneider et al.\ (2011)]{schneider}Schneider, J, Dedieu, C., 
Le Sidaner, P., Savalle, R., Zolotukhin, I., 2011, A\&A, 532, A79
\bibitem[Sestito \& Randlich (2005)]{2005A&A...442..615S} Sestito, P. \& Randlich, S., 2005, A\&A, 442, 615
\bibitem[Smalley \etal\ (2011)]{w34}Smalley B. \etal\ 2011, A\&A, 526, 130
\bibitem[Smith \etal\ (2012)]{w36}Smith A. M. S. \etal\ 2012, AJ, 143, 81
\bibitem[Southworth]{southworth}Southworth J., 2011, MNRAS, 417, 2166
\bibitem[Triaud et al.\ (2010)]{triaud10}Triaud, A. H. M. J.,  2010, A\&A, 524, 25
\bibitem{zacharias}Zacharias, N. \etal\ 2010, AJ, 139, 2184

\end{thebibliography}
\end{document}